\documentclass[12pt]{article}
\usepackage{epsfig}
\begin{document}
\title{Insights into the exotic charged states $Z_b(10610)$ and $Z_b(10650)$ from their photoproduction off nuclei}
\author{E. Ya. Paryev\\
{\it Institute for Nuclear Research of the Russian Academy of Sciences}\\
{\it Moscow, Russia}}

\renewcommand{\today}{}
\maketitle

\begin{abstract}
The possibility to study the nature of the famous charged bottomonium-like states $Z_b(10610)$ and $Z_b(10650)$,
which is by far the best known, from their inclusive photoproduction off nuclei near the kinematic threshold is investigated within the collision model based on the nuclear spectral function. The model accounts for   $Z_b(10610)^{\pm}$ and $Z_b(10650)^{\pm}$ production in direct photon--nucleon interactions as well as four different scenarios for their intrinsic configurations: compact tetraquarks, molecules of the two open-beauty mesons and two mixtures of both of them for each of $Z_b$ state. We calculate within these scenarios the absolute and relative excitation functions on $^{12}$C and $^{184}$W nuclei at photon energies of 61--90 GeV, the absolute momentum differential cross sections and ratios of them for their production off these target nuclei at laboratory polar angles of 0$^{\circ}$--5$^{\circ}$ and for photon energy of 75 GeV as well as the A-dependences of the transparency ratios for the $Z_b(10610)^{\pm}$ mesons at photon energy of 75 GeV. We show that the absolute and relative observables considered reveal distinct sensitivity to the $Z_b(10610)^{\pm}$ and $Z_b(10650)^{\pm}$ internal structures. Therefore, they might be useful for the determination of these structures from the comparison of them with the experimental data from the future high-precision experiments at the upcoming experimental facilities, such as the planned high-luminosity electron-ion colliders in the United States and China.
\end{abstract}

\newpage

\section*{1. Introduction}

\hspace{1.5cm} In the recent years, the numerous exotic heavy hadronic states, which exhibit properties that
cannot be accommodated by the traditional quark model of mesons as $q{\bar q}$ and baryons as $qqq$, have been
observed in high-energy experiments. These states are composed of four or five quarks (and antiquarks) and are named as tetraquark and pentaquark states, respectively.
In particular, various experiments have discovered, starting from the discovery of the $X(3872)$ in 2003 by the Belle Collaboration [1], a series of unconventional narrow charmonium- and bottomonium-like states with hidden charm or beauty, charged or neutral, usually referred to as $XYZ$ states as well as the hidden-charm non-strange $P_c$ and strange $P_{cs}$ pentaquark states, the doubly-charmed tetraquark $T^+_{cc}(3875)$ state, the fully charmed tetraquark $X(6900)$ state, the excited $\Omega_c$ structures and so on, as summarized in recent review papers [2--7]. These experimental discoveries have stimulated an extensive theoretical studies to understand their properties [8--16].
They have offered different exotic interpretations of these states, including compact tetraquark and pentaquark states, meson-meson and meson-baryon molecular states, hadro-quarkonium and charmonium-molecule mixtures or the product of rescattering and threshold cusp effects (see, for example, Refs. [17--20]), which are far from reaching a consensus.

Among these exotic hadrons, the first charged bottomonium-like narrow states $Z_b(10610)^{\pm}$ and $Z_b(10650)^{\pm}$
\footnote{$^)$In what follows denoted as $Z_b^{\pm}$ and $Z_b^{\prime{\pm}}$, respectively.}$^)$, which were observed by the Belle Collaboration in the mass spectra of the $\pi^{\pm}\Upsilon(nS)$ ($n=$ 1, 2, 3) and $\pi^{\pm}h_b(mP)$ ($m=$ 1, 2) pairs that are produced at the KEKB asymmetric-energy $e^+e^-$ collider in association with a single charged pion in  $\Upsilon(5S)$ decays [21], have attracted much attention as the first undoubted charged tetraquark candidates in the bottom sector. The measured masses and widths of the two states averaged over the five final states are $m_{Z_b^{\pm}}=$ (10607.2$\pm$2.0) MeV, $\Gamma_{Z_b^{\pm}}=$ (18.4$\pm$2.4) MeV and
$m_{Z_b^{\prime{\pm}}}=$ (10652.2$\pm$1.5) MeV, $\Gamma_{Z_b^{\prime{\pm}}}=$ (11.5$\pm$2.2) MeV. Later, Belle confirmed
[22] their observation [21].
The $Z_b(10610/10650)^{\pm} \to \pi^{\pm}\Upsilon(nS)$ and $Z_b(10610/10650)^{\pm} \to \pi^{\pm}h_b(mP)$ decays indicate that they are definitely tetraquark states with quark contents $b{\bar b}u{\bar d}$ and $b{\bar b}{\bar u}d$, respectively. Their existence were also confirmed by the Belle Collaboration
in the analysis of the $e^+e^- \to \Upsilon(5S) \to Z_b(10610)^{\pm}\pi^{\mp} \to [B{\bar B}^*+c.c]^{\pm}\pi^{\mp}$ and
$e^+e^- \to \Upsilon(5S) \to Z_b(10650)^{\pm}\pi^{\mp} \to [B^*{\bar B}^*]^{\pm}\pi^{\mp}$
processes [23]. In 2013, the Belle Collaboration observed [24] the neutral partner $Z_b(10610)^0$ of the $Z_b(10610)^{\pm}$ in the $\pi^0\Upsilon(2,3S)$ mass spectra from the $\Upsilon(5S) \to \pi^0\pi^0\Upsilon(2,3S)$ decays. Amplitude analysis of $e^+e^- \to \pi^+\pi^-\Upsilon(nS)$ at $\sqrt{s}=$ 10.865 GeV, performed in Ref. [22], strongly
favors $I(J^P)=1(1^+)$ isospin-spin-parity quantum number assignments for the two charged bottomonium-like $Z_b^{\pm}$ and $Z_b^{\prime{\pm}}$ states. It is worth noting that the charged charmonium-like structures near the $D{\bar D}^*$ mass threshold, $Z_c(3900)^{\pm}$, which have been observed in 2013 simultaneously by the BESIII [25] and Belle [26] Collaborations in $e^+e^-$ collisions in the ${J/\psi}\pi^{\pm}$ invariant mass distributions from the sequential process $e^+e^- \to Y(4260) \to {\pi^{\mp}}Z_c(3900)^{\pm}\to {\pi^{\mp}}({\pi^{\pm}}J/\psi) \to \pi^+\pi^-{J/\psi}$ at center-of-mass energy of 4.26 GeV, as well as those of $Z_c(4020)^{\pm}$ and $Z_c(4025)^{\pm}$ also discovered at BESIII in these collisions near the $D^*{\bar D}^*$ mass threshold in the experiments [27, 28] and [29, 30], respectively, are the charmonium analogues of the two charged bottomonium-like resonances $Z_b(10610)^{\pm}$ and $Z_b(10650)^{\pm}$ in the bottom sector. Since the measured masses of the $Z_b(10610)^{\pm}$ and $Z_b(10650)^{\pm}$ states are only a few MeV above the thresholds of the open-beauty $B{\bar B}^*$ and $B^*{\bar B}^*$ channels, which are the most dominant decay modes for them as well [23], they can be naturally interpreted as $S$-wave $B{\bar B}^*$ and $B^*{\bar B}^*$ loosely bound, resonant or virtual molecular states from the $B{\bar B}^*$ and $B^*{\bar B}^*$ interactions (see, for example, Refs. [21, 31--34] and those cited below)
\footnote{$^)$It should be pointed out that the existence of a loosely bound $S$-wave $B{\bar B}^*$ and
$B^*{\bar B}^*$ molecular states was predicted in Refs. [35, 36] before the experimental observation of the $Z_b$
resonances. A systematic exploration of the mass spectra of the hidden-bottom and double-bottom molecular tetraquark states up to $P$-wave configurations was conducted in the very recent papers [37, 38].}$^)$.
There have been other assignments of the $Z_b(10610)$ and $Z_b(10650)$ states, such as the compact tetraquark states [39, 40] -- the tightly packed states where the heavy and light quarks overlap, hadro-quarkonium states [41] -- the states consisting of a compact color-singlet heavy quarkonium and a light meson bound by hadronic interactions, quark-gluon hybrids [42] -- the states in which the heavy quark-antiquark pair is embedded in the gluon field, mixtures of the compact tetraquark states and molecular states [43--46], kinematic threshold cusp effects [47--49], etc (see also references herein below). But, despite a lot of theoretical and experimental efforts, the deep understanding of the intrinsic structures of the exotic $Z_b(10610)$ and $Z_b(10650)$ states, which is the most important issue of heavy quarkonium physics, is still lacking and further studies are needed here.

To gain further insights into the exotic $Z_b(10610)$ and $Z_b(10650)$ states and to make a definite decision about their nature, it is of foremost importance to study their photoproduction on nuclei at energies close to the thresholds for their production off a free nucleon. This has the advantage compared to the hadronic collisions  that the interpretation of data from such experiments is less ambiguous owing to a negligible strength of initial-state photon interaction and since their production proceeds through a few channels in a cleaner environment - in static cold nuclear matter whose density is sufficiently well known.
The study of the photoproduction of $Z_b(10610)^{\pm}$ and $Z_b(10650)^{\pm}$ states off a proton target has been carried out previously in Refs. [50--52]. Their exclusive and inclusive production cross sections from pion exchange have been predicted. The production and absorption of $Z_b(10610)$ and $Z_b(10650)$ states in a hadronic medium via the processes ${\bar B}B \to {\pi}Z_b(10610)$, ${\bar B}^*B \to {\pi}Z_b(10610)$, ${\bar B}^*B^* \to {\pi}Z_b(10610)$ and
${\bar B}^*B \to {\pi}Z_b(10650)$, ${\bar B}^*B^* \to {\pi}Z_b(10650)$
and the corresponding inverses reactions was studied in Ref. [53]. The absorption cross sections were found here to
be greater than the production ones, but still comparable with them. This fact may give a chance of essential survival probability of these states in high-energy heavy-ion collisions. The production of the hidden-bottom hadronic molecules
$Z_b(10610)$ and $Z_b(10650)$ at the $Z$ factory of future Circular Electron Positron Collider (CEPC) [54, 55] has been theoretically studied in Ref. [56]. Additionally, it could be also investigated experimentally at the Future Circular Collider (FCC) running in the $e^+e^-$ mode (FCC-ee) at the $Z$-pole [57--61].
In this work we present the detailed predictions of absolute and relative observables for $Z_b(10610)^{\pm}$ and
$Z_b(10650)^{\pm}$ photoproduction off nuclei at threshold energies obtained in the framework of the collision model based on the nuclear spectral function within four different scenarios for their internal structures. The predictions can be confronted to the experimental data also from the future measurements at electron-ion colliders
EIC [62, 63] and EicC [64, 65] to distinguish between these scenarios.

\section*{2. Direct $Z_b(10610)^{\pm}$ and $Z_b(10650)^{\pm}$ photoproduction mechanisms: cross sections and their ratios}

\hspace{1.5cm} Direct production of the charged components $Z_b(10610)^{\pm}$ and $Z_b(10650)^{\pm}$ off nuclei in the near-threshold photon beam energy region $E_{\gamma} \le 90$ GeV
\footnote{$^)$Which corresponds to the center-of-mass energies $W$ of the photon-proton system $W \le 13.0$ GeV. At these energies the $Z_b(10610)^{\pm}$ and $Z_b(10650)^{\pm}$ elementary production processes with one additional pion in the final states can be ignored [51, 52] and where they can be observed in the ${\gamma}p$ and ${\gamma}A$ reactions at the proposed electron-ion colliders EIC [62, 63] and EicC [64, 65].}$^)$
can occur in the following ${\gamma}p$ and ${\gamma}n$ elementary processes [50--52]:
\begin{equation}
{\gamma}+p \to Z_b(10610/10650)^++n,
\end{equation}
\begin{equation}
{\gamma}+n \to Z_b(10610/10650)^-+p
\end{equation}
with the lowest free production threshold energies ($\approx$ 70.5797 and 71.1346 GeV, respectively).
The $Z_b(10610/10650)^{\pm}$ mesons and nucleons, produced in these processes, are sufficiently energetic.
Thus, for example, the kinematically allowed $Z_b(10610)^+$ meson and final neutron laboratory momenta in the direct process (1), proceeding on the free target proton at rest, vary within the momentum ranges of 61.700--72.417 GeV/c and 2.583--13.300 GeV/c, respectively, at photon energy of $E_{\gamma}=75$ GeV. The kinematical characteristics of another particles, produced in the reactions (1), (2), are similar to those mentioned above.
Since the medium effects are expected to be reduced for high momenta, we will neglect the medium modifications of the final high-momentum hadrons in the case when the reactions (1), (2) proceed on the nucleons embedded in a nuclear matter.

Then, neglecting the distortion of the incident photon
\footnote{$^)$Accounting for this distortion leads to the reduction of the total cross sections (3), (4) for the production of the $Z_b^{\pm}$ mesons on carbon and tungsten target nuclei, as our calculations have shown, only by a small fractions about of 0.2 and 0.8\%, respectively.}$^)$
and describing the $Z_b(10610)^{\pm}$ absorption by intranuclear nucleons by the absorption cross sections $\sigma_{Z_b^{\pm}p}$ and $\sigma_{Z_b^{\pm}n}$, we represent the total cross sections for the production of $Z_b^{\pm}$ mesons on nuclei from the direct photon--induced reactions (1), (2) as follows [66]:
\begin{equation}
\sigma_{{\gamma}A\to Z_b^+X}^{({\rm dir})}(E_{\gamma})=I_{V}[A,\rho_p,\sigma_{Z_b^+p},\sigma_{Z_b^+n}]
\left<\sigma_{{\gamma}p \to Z_b^+n}(E_{\gamma})\right>_A,
\end{equation}
\begin{equation}
\sigma_{{\gamma}A\to Z_{b}^-X}^{({\rm dir})}(E_{\gamma})=I_{V}[A,\rho_n,\sigma_{Z_{b}^-p},\sigma_{Z_{b}^-n}]
\left<\sigma_{{\gamma}n \to Z_{b}^-p}(E_{\gamma})\right>_A;
\end{equation}
where
\begin{equation}
I_{V}[A,\rho_{p(n)},\sigma_{Z_bp},\sigma_{Z_bn}]=2{\pi}\int\limits_{0}^{R}r_{\bot}dr_{\bot}
\int\limits_{-\sqrt{R^2-r_{\bot}^2}}^{\sqrt{R^2-r_{\bot}^2}}dz
\rho_{p(n)}(\sqrt{r_{\bot}^2+z^2})\times
\end{equation}
$$\times
\exp{\left[-\sigma_{Z_bp}\int\limits_{z}^{\sqrt{R^2-r_{\bot}^2}}\rho_p(\sqrt{r_{\bot}^2+x^2})dx-
\sigma_{Z_bn}\int\limits_{z}^{\sqrt{R^2-r_{\bot}^2}}\rho_n(\sqrt{r_{\bot}^2+x^2})dx
\right]}.
$$
\begin{equation}
\left<\sigma_{{\gamma}p \to {Z_b^+}n({\gamma}n \to {Z_b^-}p)}(E_{\gamma})\right>_A=
\int\int
P_A({\bf p}_t,E)d{\bf p}_tdE
\sigma_{{\gamma}p \to {Z_b^+}n({\gamma}n \to {Z_b^-}p)}(\sqrt{s^*}).
\end{equation}
Here, $\rho_p(r)$ and  $\rho_n(r)$ ($r$ is the distance from the nucleus center) are normalized to the numbers of protons $Z$ and neutrons $N$ the local proton and neutron densities of the target nucleus with mass number $A$ ($A=Z+N$), having mass $M_A$ and radius $R$. And $\sigma_{{\gamma}p \to {Z_b^+}n}(\sqrt{s^*})$, $\sigma_{{\gamma}n \to {Z_b^-}p}(\sqrt{s^*})$ are "in-medium" total cross sections for the production of $Z_b(10610)^{\pm}$ mesons in reactions (1), (2) at the in-medium ${\gamma}p$ center-of-mass energy $\sqrt{s^*}$
\footnote{$^)$We neglect the difference between proton ($m_p$) and neutron ($m_n$) masses.}$^)$. They are averaged over target nucleon binding and Fermi motion encoded in the spectral function $P_A({\bf p}_t,E)$ of target nucleus
\footnote{$^)$The specific information about this quantity, used in our calculations, is given in Ref. [67].}$^)$.
The total cross sections for the production of $Z_b(10650)^{+}$ and $Z_b(10650)^{-}$ mesons on nuclei from the processes (1) and (2) can be expressed by Eqs. (3) and (4), respectively, in which one needs to make the substitutions: $Z_b^{+} \to Z_b^{\prime{+}}$ and $Z_b^{-} \to Z_b^{\prime{-}}$.

As before in Ref. [66], we assume that the "in-medium" elementary cross sections for $Z_b^{\pm}$, $Z_b^{\prime{\pm}}$ production in reactions (1), (2) are equivalent to the respective vacuum cross sections, in which the center-of-mass energy squared $s$ of the free space photon-proton system for given photon laboratory energy $E_{\gamma}$ and momentum ${\bf p}_{\gamma}$, presented by the formula
\begin{equation}
s=s(E_{\gamma})=W^2=(E_{\gamma}+m_p)^2-{\bf p}_{\gamma}^2=m_p^2+2m_pE_{\gamma},
\end{equation}
is replaced by the in-medium expression
\begin{equation}
  s^*=(E_{\gamma}+E_t)^2-({\bf p}_{\gamma}+{\bf p}_t)^2,
\end{equation}
\begin{equation}
   E_t=M_A-\sqrt{(-{\bf p}_t)^2+(M_{A}-m_{p}+E)^{2}}.
\end{equation}
Here, $E_t$, ${\bf p}_{t}$ and $E$ are the total energy, internal momentum and binding energy of the
struck target nucleons involved in the collision processes (1), (2).

For the free total cross sections of the photon-induced reaction channels (1), (2) no data are available
presently at the considered photon energies $E_{\gamma} \le $ 90 GeV. Therefore, we have to rely on some theoretical predictions and estimates for them, existing in the literature at these energies. For the free total cross sections $\sigma_{{\gamma}p \to Z_b^+n}(\sqrt{s})$
and $\sigma_{{\gamma}p \to Z_b^{\prime{+}}n}(\sqrt{s})$ of the reactions (1) in the considered photon energy range we have used the following parametrizations of the results of calculations of these cross sections here within the vector-meson-dominance model [51, 52]
\footnote{$^)$In this model, an initial real photon fluctuates into a virtual vector meson $\Upsilon(1S,2S,3S)$, which
subsequently scatters inelastically off the target proton through a positively charged pion exchange to the final states
consisting of the $Z_b$, $Z_b^{\prime{+}}$ and neutrons. To calculate the photoproduction cross sections of these processes the measured branching ratios of $Z_b^+$, $Z_b^{\prime{+}}$ states to $\Upsilon(1S,2S,3S){\pi}$ and the
established quarkonium phenomenology are used without making any assumptions about their specific internal structure.}$^)$:
\begin{equation}
\sigma_{{\gamma}p \to {Z_b^+}n}(\sqrt{s})=3.95\left(1-\frac{s_{\rm th}}{s}\right)^{0.302}~[\rm nb],
\end{equation}
\begin{equation}
\sigma_{{\gamma}p \to Z_b^{\prime{+}}n}(\sqrt{s})=1.065\left(1-\frac{{\tilde s}_{\rm th}}{s}\right)^{0.362}
~[\rm nb],
\end{equation}
where
\begin{equation}
s_{\rm th}=(m_{Z_b}+m_{n})^2=(11.54675~{\rm GeV})^2, \,\,
{\tilde s}_{\rm th}=(m_{Z_b^{\prime}}+m_{n})^2=(11.59175~{\rm GeV})^2.
\end{equation}
Here, $m_{Z_b}$ and $m_{Z_b^{\prime}}$ are the free space $Z_b$ and $Z_b^{\prime}$ masses.
In line with the "isospin considerations" of the processes (1), (2), it is natural to assume that the following relations among the total cross sections of the $Z_b(10610)^{\pm}$ and $Z_b(10650)^{\pm}$ production channels exist:
\begin{equation}
\sigma_{{\gamma}n \to {Z_b^-}p}(\sqrt{s})\approx\sigma_{{\gamma}p \to {Z_b^+}n}(\sqrt{s}),\,\,
\sigma_{{\gamma}n \to Z_b^{\prime{-}}p}(\sqrt{s})\approx\sigma_{{\gamma}p \to Z_b^{\prime{+}}n}(\sqrt{s}).
\end{equation}
It is worth mentioning that the approach [51, 52] predicts that the total cross sections of the
${\gamma}p \to {Z_b^+}n$ and ${\gamma}p \to {Z_b^{\prime{+}}}n$ reactions are of about 1.8--2.9 nb and 0.4--0.7 nb
for the near-threshold ${\gamma}p$ c.m. energies $W=$ 12--15 GeV
\footnote{$^)$It should be noticed that these values are considerably greater (for example, by about of three orders of magnitude at c.m. energy $W=$ 15 GeV) than the results for these cross sections obtained in the $\pi$ exchange model [50].}$^)$,
which are well within reach of the future electron-ion colliders.
Evidently, to put a strong constraints on the ${Z_b}$ and ${Z_b^{\prime}}$ internal structures using their absolute yields from nuclei one needs to know precisely their proton-target production cross sections.
Meanwhile, the following tells us in favor that these cross sections are realistic enough.
It is expected that the yields of the exotic bottom states are one-two orders of magnitude lower than those of their counterparts in the charm sector (cf. Refs. [64, 65, 68]) at the same excess energies above the respective thresholds. The results on the total cross sections of the reactions ${\gamma}p \to Z_c(3900)^+n$ and ${\gamma}p \to Z_b(10610)^+n$,
${\gamma}p \to Z_b(10650)^+n$, obtained within the same model [51, 52], support this expectation. Thus, for example, for the excess energy of about 1.2 GeV, corresponding to the photon-proton c.m. energies of about 6 and 13 GeV in the former and latter reactions, this model predicts for the above total cross sections the values of the order of 25 nb and 2.5 nb, 0.5 nb. They lead to the respective ratios of them of about 10 and 50. In addition, the approach [51, 52] predicts that the total cross section of the ${\gamma}p \to {Z_c(3900)^+}n$ reaction is of about 1.4 nb for the ${\gamma}p$ c.m. energy $W=$ 14 GeV. This value is comparable with the upper limit for the proton-target $Z_c(3900)$ exclusive photoproduction cross section of $\sim$ 1 nb measured by the COMPASS Collaboration at an average energy of $<W>=$ 13.8 GeV, once the relevant branching ratio is taken into account [69]. These facts together with the fact that
the relative observables (transparency ratios $S_A$ and $T_A$) considered in the present work are less sensitive than absolute cross sections themselves to the theoretical uncertainties associated mainly with the experimentally unknown total cross sections of the elementary reactions (1), (2) due to their essential cancelation in them
are a convincing advantage of the predicted in [51, 52] $Z_b$ and ${Z_b^{\prime}}$ near-threshold photoproduction cross sections on a proton target compared to those from Ref. [50]. They reduce model dependence and may give a confidence to us that the results (both absolute and relative) obtained in the present work (see below), employing these cross sections, may be an important tool to provide further insight into the $Z_b$ and ${Z_b^{\prime}}$ inner structures.

The local proton and neutron densities, adopted in the calculations of the quantities\\
$I_{V}[A,\rho_{p(n)},\sigma_{Z_bp},\sigma_{Z_bn}]$ entering into Eqs. (3) and (4), for the target nuclei $^{12}_{6}$C, $^{27}_{13}$Al, $^{40}_{20}$Ca, $^{63}_{29}$Cu, $^{93}_{41}$Nb, $^{112}_{50}$Sn, $^{184}_{74}$W, $^{208}_{82}$Pb and $^{238}_{92}$U considered in the present work are given in Ref. [66]. As in Ref. [66], for medium-weight $^{93}_{41}$Nb, $^{112}_{50}$Sn and heavy $^{184}_{74}$W, $^{208}_{82}$Pb, $^{238}_{92}$U target nuclei we use the neutron density $\rho_n(r)$ in the 'skin' form.

To evaluate the rates of the $Z_b^{\pm}$ (${Z_b^{\prime{\pm}}}$) photoproduction on nuclei in different scenarios for their intrinsic structures, we have to specify the effective input $Z_b^{\pm}$(${Z_b^{\prime{\pm}}}$)--nucleon absorption cross sections $\sigma_{Z_b^{\pm}({Z_b^{\prime{\pm}}})p}$ and $\sigma_{Z_b^{\pm}({Z_b^{\prime{\pm}}})n}$ in these scenarios. They determine the numbers (5) of the target nucleons participating in the direct processes (1), (2). With this aim, in our study we consider four different popular scenarios for the $Z_b^{\pm}$ and ${Z_b^{\prime{\pm}}}$ [20]: i) compact, $\sim$ 1 fm, diquark-antidiquark tetraquark bound states: relatively tightly bound pairs $[bu]$ and $[{\bar b}{\bar d}]$ for the $Z_b^+(Z_b^{\prime{+}})$ and $[bd]$ and $[{\bar b}{\bar u}]$ ones for their charge conjugate $Z_b^-(Z_b^{\prime{-}})$, which interact by the gluonic color force [39, 40, 70--74], ii) $B{\bar B}^*$ and $B^*{\bar B}^*$ hadronic molecules: $S$-wave molecular states (bound, virtual or resonant) formed by pairs of heavy bottom and antibottom mesons [17, 21, 31--34, 75--94]
\footnote{$^)$The measured masses of the $Z_b(10610)^{\pm}$ and $Z_b(10650)^{\pm}$ states are only a few MeV above the thresholds of the open-beauty $B{\bar B}^*$ and $B^*{\bar B}^*$ channels. Therefore, it is natural to suppose that these are a good candidates for either a $B{\bar B}^*$ and $B^*{\bar B}^*$ virtual states below the relevant lowest thresholds or an above-thresholds molecular resonances on the second unphysical Riemann sheets of the scattering amplitudes (see, for example, Refs. [32--34, 77, 83, 85]).}$^)$,
and iii, iv) hybrid states: states in which each of the $Z_b^{\pm}$ and $Z_b^{\prime{\pm}}$ is considered as two different  mixtures of the compact diquark-antidiquark and molecular components [43--46].

The observed masses of the two bottomonium-like structures $Z_b^{\pm}$ and ${Z_b^{\prime{\pm}}}$ are only about 3 MeV
above the respective $B{\bar B}^*$ and $B^*{\bar B}^*$ thresholds [21]. This suggests a parallel with the famous $X(3872)$ resonance whose mass is almost exactly at the $D^0{\bar D}^{*0}$ (${\bar D}^0D^{*0}$) threshold [95]. Therefore, interpreting the $Z_b^{\pm}$ (${Z_b^{\prime{\pm}}}$) as a compact tetraquarks, it is natural to assume for the high-momentum $Z_b^{\pm}$ (${Z_b^{\prime{\pm}}}$)--proton (neutron) absorption cross sections $\sigma_{{Z_b^{\pm}}p(n)}^{{\rm 4{q}}}$ ($\sigma_{{Z_b^{\prime{\pm}}}p(n)}^{{\rm 4{q}}}$) in this picture the same values as that for the absorption cross section of the $X(3872)$ resonance with similar momenta. Considering it as a compact tetraquark state with radius $r_{4q}=0.65$ fm and approximating the latter cross section by a geometrical cross section
$\sigma^{\rm geo}_{4q}=\pi{r_{4q}^2}$, we have $\sigma_{{Z_b^{\pm}}p(n)}^{{\rm 4{q}}}=\sigma_{{Z_b^{\prime{\pm}}}p(n)}^{{\rm 4{q}}}=13.3$ mb [66].

Within the "pure" hadronic molecular interpretation of the $Z_b(10610)^{\pm}$ and $Z_b(10650)^{\pm}$, their wave functions take the following charge-conjugation forms [33, 36, 76, 78, 81, 85]:
\begin{equation}
|Z_b(10610)^+>_{\rm mol}=\frac{1}{\sqrt{2}}\left(|B^+{\bar B}^{*0}>+|B^{*+}{\bar B}^0>\right),
\end{equation}
\begin{equation}
|Z_b(10610)^->_{\rm mol}=\frac{1}{\sqrt{2}}\left(|B^-B^{*0}>+|B^{*-}B^0>\right)
\end{equation}
and
\begin{equation}
|Z_b(10650)^+>_{\rm mol}=|B^{*+}{\bar B}^{*0}>,
\end{equation}
\begin{equation}
|Z_b(10650)^->_{\rm mol}=|B^{*-}B^{*0}>.
\end{equation}

Following [66], we will take the point of quasi-free approximation
view that the $Z_b^{\pm}$ and ${Z_b^{\prime{\pm}}}$ states dissociate while one of their constituent bottom mesons flying with an average laboratory momentum $\sim$ 35 GeV/c (see above) scatters (elastically or inelastically) with the target nucleon (proton or neutron) and the other bottom meson is a spectator.
It should be noted out that this suggestion is sufficiently well justified only when the average distance between the constituents in the hadronic molecule is much larger than a typical radius of strong interactions (a few fermi). In this case the constituents can be considered as on-shell individual particles flying together and interactions between them can be ignored [96]. But in our present case of $Z_b^{\pm}$ and $Z_b^{\prime{\pm}}$ states,
an excitation energy $\sim$ 2.7 MeV over the respective thresholds pushes this picture to its limit of validity. Thus, adopting the expression $r_{Z_b(Z_b^{\prime})}=1/\sqrt{4\mu_0|\delta_{Z_b(Z_b^{\prime})}|}$ for the r.m.s. size $r_{Z_b(Z_b^{\prime})}$ of the intermeson distance [66], where $\mu_0$ is the $B{\bar B}^*$ ($B^*{\bar B}^*$) reduced mass and $\delta_{Z_b(Z_b^{\prime})}$ is "the binding energy" $\delta_{Z_b}=m_{B}+m_{{\bar B}^*}-m_{Z_b}$ $(\delta_{Z_b^{\prime}}=m_{B^*}+m_{{\bar B}^*}-m_{Z_b^{\prime}})$ of the molecule, and taking $\delta_{Z_b}=\delta_{Z_b^{\prime}}=-2.7$ MeV, we find that the bottom mesons in the $Z_b$ and $Z_b^{\prime}$ have an insignificant r.m.s. separations: $r_{Z_b}=r_{Z_b^{\prime}}=1.2$ fm. This size scale is small and it is close to the range, where the bottom mesons as a spatially extended objects with mean spatial sizes $\sim$ 0.5 fm [97] start to overlap and possibly cannot be considered as individual particles. This means that the $Z_b^{\pm}$ and $Z_b^{\prime{\pm}}$ cannot be reliable described as molecular states and it may not be correct to treat their interactions with target nucleons in the manner adopted in the present work. The situation could be improved only at the price of making the sizes of open bottom mesons much smaller than what they are assumed to be. But in view of the absence of such possibility and a reliable theory, for an estimate of the $Z_b^{\pm}(Z_b^{\prime{\pm}})N$ absorption cross sections in molecular picture, we will nevertheless follow this treatment.
Then, assuming that the total cross sections of the free high-momentum ${\bar B}^{*0}N$, $B^{*0}N$ and $B^{*+}N$,  $B^{*-}N$ interactions are the same as those for the ${\bar B}^{0}N$, $B^{0}N$ and $B^{+}N$,  $B^{-}N$
ones by analogy with the open charm meson--nucleon interactions [98--100], we can estimate the cross sections for $Z_b^{\pm}(Z_b^{\prime{\pm}})$ absorption in the molecular context, $\sigma_{{Z_b^{\pm}}p(n)}^{\rm mol}(\sigma_{{Z_b^{\prime{\pm}}}p(n)}^{\rm mol})$, as:
\begin{equation}
\sigma_{{Z_b^+}p}^{\rm mol}=\sigma_{Z_b^{\prime+}p}^{\rm mol}\approx\sigma_{B^+p}^{\rm el}+\sigma_{B^+p}^{\rm in}+\sigma_{{\bar B}^0p}^{\rm el}+\sigma_{{\bar B}^0p}^{\rm in},
\end{equation}
\begin{equation}
\sigma_{{Z_b^+}n}^{\rm mol}=\sigma_{Z_b^{\prime+}n}^{\rm mol}\approx\sigma_{B^+n}^{\rm el}+\sigma_{B^+n}^{\rm in}+\sigma_{{\bar B}^0n}^{\rm el}+\sigma_{{\bar B}^0n}^{\rm in}+\sigma_{{\bar B}^0n \to B^-p}+
\sigma_{B^+n \to B^0p};
\end{equation}
\begin{equation}
\sigma_{{Z_b^-}p}^{\rm mol}=\sigma_{Z_b^{\prime-}p}^{\rm mol}\approx\sigma_{B^-p}^{\rm el}+\sigma_{B^-p}^{\rm in}+\sigma_{B^0p}^{\rm el}+\sigma_{B^0p}^{\rm in}+\sigma_{B^-p \to {\bar B}^0n}+\sigma_{B^0p \to B^+n},
\end{equation}
\begin{equation}
\sigma_{{Z_b^-}n}^{\rm mol}=\sigma_{Z_b^{\prime-}n}^{\rm mol}\approx\sigma_{B^-n}^{\rm el}+\sigma_{B^-n}^{\rm in}+\sigma_{B^0n}^{\rm el}+
\sigma_{B^0n}^{\rm in}.
\end{equation}
Here, $\sigma_{B^+p(n)}^{\rm el(in)}$ and $\sigma_{B^-p(n)}^{\rm el(in)}$ as well as $\sigma_{{\bar B}^0p(n)}^{\rm el(in)}$ and $\sigma_{B^0p(n)}^{\rm el(in)}$ are the elastic (inelastic) cross sections of the free
$B^+p$($B^+n$) and $B^-p$($B^-n$) as well as ${\bar B}^0p$(${\bar B}^0n$) and $B^0p$($B^0n$)
interactions, respectively. And, $\sigma_{{\bar B}^0n \to B^-p}$, $\sigma_{B^+n \to B^0p}$,
$\sigma_{B^-p \to {\bar B}^0n}$ and $\sigma_{B^0p \to B^+n}$ are the total cross sections of the free charge-exchange
reactions ${\bar B}^0n \to B^-p$, $B^+n \to B^0p$, $B^-p \to {\bar B}^0n$ and $B^0p \to B^+n$, correspondingly.
No data for the elastic, inelastic and charge exchange cross sections of open beauty $B$ and ${\bar B}$ meson interactions with nucleons are available. In order to estimate them, we use the analogy with open charm $D$ and ${\bar D}$ meson interactions with nucleons in terms of their quark contents. In the constituent quark model we get for the open beauty mesons $B^+=|{\bar b}u>$, $B^0=|{\bar b}d>$, $B^-=|b{\bar u}>$, ${\bar B}^0=|b{\bar d}>$ and for the open charm mesons $D^+=|c{\bar d}>$, $D^0=|c{\bar u}>$, $D^-=|{\bar c}d>$, ${\bar D}^0=|{\bar c}u>$. Due to the large masses of
$b({\bar b})$ and $c({\bar c})$ quarks, they can be considered as being frozen in the ${\bar B}(B)$ and $D({\bar D})$
mesons. And interactions of these mesons with nucleons, composed from a light quarks $u$ and $d$, may be regarded as being controlled mainly by their light quark (or antiquark) contents and not by their heavy quarks contents [101--103]. As a consequence, one can expect that the $B^+N$ interaction is equivalent to the ${\bar D}^0N$ one. Analogously, one can assume that the $B^0N$, $B^-N$ and ${\bar B}^0N$ interactions are equivalent, respectively, to the $D^-N$, $D^0N$ and $D^+N$ ones
\footnote{$^)$It should be pointed out that these assumptions are supported by the fact that the modifications of $D(\bar D)$ and $B({\bar B})$ mesons in cold nuclear matter are predicted to be similar [101, 104--106].}$^)$.
Therefore, there are the following relations between the elastic, inelastic and charge exchange cross sections of open beauty meson--nucleon and open charm meson--nucleon interactions: $\sigma_{B^+p(n)}^{\rm el(in)}=\sigma_{{\bar D}^0p(n)}^{\rm el(in)}$,
$\sigma_{B^-p(n)}^{\rm el(in)}=\sigma_{D^0p(n)}^{\rm el(in)}$,
$\sigma_{{\bar B}^0p(n)}^{\rm el(in)}=\sigma_{D^+p(n)}^{\rm el(in)}$,
$\sigma_{B^0p(n)}^{\rm el(in)}=\sigma_{D^-p(n)}^{\rm el(in)}$ and $\sigma_{{\bar B}^0n \to B^-p}=\sigma_{D^+n \to D^0p}$, $\sigma_{B^+n \to B^0p}=\sigma_{{\bar D}^0n \to D^-p}$, $\sigma_{B^-p \to {\bar B}^0n}=\sigma_{D^0p \to D^+n}$,
$\sigma_{B^0p \to B^+n}=\sigma_{D^-p \to {\bar D}^0n}$. For the latter cross sections in our calculations we adopt
the following constants which are relevant to the momentum regime above of 1 GeV/c of interest:
$\sigma_{D^+p(n)}^{\rm el}=\sigma_{D^-p(n)}^{\rm el}=\sigma_{D^0p(n)}^{\rm el}=\sigma_{{\bar D}^0p(n)}^{\rm el}=10$ mb,
$\sigma_{D^+p(n)}^{\rm in}=\sigma_{D^0p(n)}^{\rm in}=10$ mb, $\sigma_{D^-p(n)}^{\rm in}=\sigma_{{\bar D}^0p(n)}^{\rm in}=0$, $\sigma_{{\bar D}^0n \to D^-p}=\sigma_{D^+n \to D^0p}=\sigma_{D^-p \to {\bar D}^0n}=\sigma_{D^0p \to D^+n}=12$ mb [98--100].
Using these values, we get that
$\sigma_{{Z_b^+}p}^{\rm mol}=\sigma_{Z_b^{\prime+}p}^{\rm mol}=30$ mb,
$\sigma_{{Z_b^+}n}^{\rm mol}=\sigma_{Z_b^{\prime+}n}^{\rm mol}=54$ mb and
$\sigma_{{Z_b^-}p}^{\rm mol}=\sigma_{Z_b^{\prime-}p}^{\rm mol}=54$ mb,
$\sigma_{{Z_b^-}n}^{\rm mol}=\sigma_{Z_b^{\prime-}n}^{\rm mol}=30$ mb.

In the hybrid scenario, it is assumed that the $Z_b(10610)^{\pm}$ and $Z_b(10650)^{\pm}$
wave functions are a linear superpositions of the compact four-quark (or 4$q$) and molecular components [43--46]:
\begin{equation}
|Z_b(10610)^+>_{\rm hyb}=\alpha|b{\bar b}u{\bar d}>+\frac{\beta}{\sqrt{2}}\left(|B^+{\bar B}^{*0}>+|B^{*+}{\bar B}^0>\right),
\end{equation}
\begin{equation}
|Z_b(10610)^->_{\rm hyb}=\alpha|b{\bar b}{\bar u}d>+\frac{\beta}{\sqrt{2}}\left(|B^-B^{*0}>+|B^{*-}B^0>\right)
\end{equation}
and
\begin{equation}
|Z_b(10650)^+>_{\rm hyb}=\alpha|b{\bar b}u{\bar d}>+{\beta}|B^{*+}{\bar B}^{*0}>,
\end{equation}
\begin{equation}
|Z_b(10650)^->_{\rm hyb}=\alpha|b{\bar b}{\bar u}d>+{\beta}|B^{*-}B^{*0}>.
\end{equation}
Here, the amplitudes $\alpha$ and $\beta$ squared, $\alpha^2$ and $\beta^2$, represent the probabilities to find a relevant "elementary" compact non-molecular and two-body molecular hadronic components in the $Z_b(10610/10650)^{\pm}$ resonances, respectively. They are normalized as
\begin{equation}
\alpha^2+\beta^2=1
\end{equation}
and are named in the literature [107, 108], correspondingly, as the "elementariness" (or the "elementarity" ) $Z$ and "compositeness" $X$. The limiting case of $\alpha^2=0$, $\beta^2=1$ corresponds to the pure molecular interpretation of the $Z_b(10610)^{\pm}$ and $Z_b(10650)^{\pm}$ states, while the case of $\alpha^2=1$, $\beta^2=0$ refers to their pure compact four-quark treatment. With a combined analysis of data on $\Upsilon(5S) \to h_b(1P,2P){\pi^+}\pi^-$ and $\Upsilon(5S) \to B^*{\bar B}^*\pi$ in an effective field theory approach the authors of Ref. [43] determine
that the probability of finding the compact components in $Z_b$ and $Z_b^{\prime}$ states may be as large as about 40\%.
The values of the compositeness $X$ of around 0.6 in $Z_b(10610)$ and $Z_b(10650)$ resonances were obtained in Ref. [44] from their pole positions. We see that they are similar to those in Ref [43]. The authors of Ref. [45] found from data analysis that the compositeness $X$ can range from about 0.4 up to 1 for the molecular $B{\bar B}^*$($B^*{\bar B}^*$) component in the resonance $Z_b(10610)$($Z_b(10650)$). On the other hand, the results of Ref. [46] indicate that the
$Z_b(10610)$ and $Z_b(10650)$ states can be explained as hadronic molecules slightly mixing with diquark-antidiquark
states
\footnote{$^)$For both the $Z_b(10610)$ and $Z_b(10650)$ states, the molecular components are about 97\% compared
to around 3\% diquark-antidiquark state components. Evidently, this case corresponds practically to the pure molecular interpretation of the $Z_b(10610)$ and $Z_b(10650)$ resonances.}$^)$.
In line with the above, following Refs. [43--45], we will assume that in the hybrid scenario the $Z_b(10610)^{\pm}$ and $Z_b(10650)^{\pm}$ wave functions (22)--(25) contain 50\% of the compact tetraquark 4$q$ components and 50\% of the molecular components ($\alpha^2=\beta^2=0.5$). In addition, to extend the range of applicability of our model and to see the sensitivity of the $Z_b^{\pm}$ and $Z_b^{\prime{\pm}}$ production cross sections from the direct processes (1), (2) to the non-molecular and molecular probabilities of the $Z_b^{\pm}$ and $Z_b^{\prime{\pm}}$ we will further adopt in calculations, following Ref. [45], another additional option for them, namely: 25\% and 75\% ($\alpha^2=0.25$ and $\beta^2=0.75$). These two options for the non-molecular and molecular probabilities of the $Z_b^{\pm}$ and $Z_b^{\prime{\pm}}$ cover the bulk of theoretical information available presently in this field.
In the hybrid treatment (22)--(25) of the $Z_b(10610)^{\pm}$ and $Z_b(10650)^{\pm}$, we can represent  the $Z_b^{\pm}$($Z_b^{\prime{\pm}}$)--proton (neutron) absorption cross sections $\sigma_{{Z_b^{\pm}}p(n)}^{\rm hyb}$($\sigma_{Z_b^{\prime{\pm}}p(n)}^{\rm hyb}$) in the following incoherent probability-weighted sum [109]:
\begin{equation}
\sigma_{Z_b^{\pm}p(n)}^{\rm hyb}(\sigma_{Z_b^{\prime{\pm}}p(n)}^{\rm hyb})=\alpha^2\sigma_{Z_b^{\pm}p(n)}^{\rm 4{q}}(\sigma_{Z_b^{\prime{\pm}}p(n)}^{\rm 4q})+\beta^2\sigma_{Z_b^{\pm}p(n)}^{\rm mol}(\sigma_{Z_b^{\prime{\pm}}p(n)}^{\rm mol}).
\end{equation}
According to the above, we set $\sigma_{Z_b^{\pm}p(n)}^{\rm 4{q}}=\sigma_{Z_b^{\prime{\pm}}p(n)}^{\rm 4{q}}=13.3$ mb and $\sigma_{Z_b^+p(n)}^{\rm mol}=\sigma_{Z_b^{\prime+}p(n)}^{\rm mol}=30~(54)$
mb, $\sigma_{Z_b^-p(n)}^{\rm mol}=\sigma_{Z_b^{\prime-}p(n)}^{\rm mol}=54~(30)$ mb. With these values, the $\sigma_{Z_b^{\pm}p(n)}^{\rm hyb}$ ($\sigma_{Z_b^{\prime{\pm}}p(n)}^{\rm hyb}$)
absorption cross sections (27) are $\sigma_{Z_b^{+}p(n)}^{\rm hyb}=\sigma_{Z_b^{\prime+}p(n)}^{\rm hyb}=21.65~(33.65)$ mb and $\sigma_{Z_b^{-}p(n)}^{\rm hyb}=\sigma_{Z_b^{\prime-}p(n)}^{\rm hyb}=33.65~(21.65)$ mb for the non-molecular and molecular probabilities of the $Z_b^{\pm}$ and $Z_b^{\prime{\pm}}$ 50\% and 50\%. For these probabilities of
25\% and 75\% the above "hybrid" cross sections are $\sigma_{Z_b^{+}p(n)}^{\rm hyb}=\sigma_{Z_b^{\prime+}p(n)}^{\rm hyb}=25.825~(43.825)$ mb and $\sigma_{Z_b^{-}p(n)}^{\rm hyb}=\sigma_{Z_b^{\prime-}p(n)}^{\rm hyb}=43.825~(25.825)$ mb.
Now, we summarize here the results obtained above for the $Z_b^{\pm}$($Z_b^{\prime{\pm}}$)--proton (neutron) absorption cross sections $\sigma_{{Z_b^{\pm}}p(n)}(\sigma_{{Z_b^{\prime{\pm}}}p(n)})$ in the adopted scenarios for the $Z_b^{\pm}$ and $Z_b^{\prime{\pm}}$:
\begin{equation}
\sigma_{{Z_b^+}p}(\sigma_{{Z_b^{\prime+}}p})=\left\{
\begin{array}{llll}
	13.3~{\rm mb}
	&\mbox{for tetraquark (4q) state}, \\
	&\\
    21.65~{\rm mb}
	&\mbox{for hybrid state (50\%,50\%)},\\
	&\\
    25.825~{\rm mb}
    &\mbox{for hybrid state (25\%,75\%)},\\
	&\\
    30~{\rm mb}
	&\mbox{for $B{\bar B}^*(B^*{\bar B}^*)$ molecule};
\end{array}
\right.	
\end{equation}
\begin{equation}
\sigma_{{Z_b^+}n}(\sigma_{{Z_b^{\prime+}}n})=\left\{
\begin{array}{llll}
	13.3~{\rm mb}
	&\mbox{for tetraquark (4q) state}, \\
	&\\
    33.65~{\rm mb}
	&\mbox{for hybrid state (50\%,50\%)},\\
	&\\
    43.825~{\rm mb}
    &\mbox{for hybrid state (25\%,75\%)},\\
	&\\
    54~{\rm mb}
	&\mbox{for $B{\bar B}^*(B^*{\bar B}^*)$ molecule}
\end{array}
\right.	
\end{equation}
and
\begin{equation}
\sigma_{{Z_b^-}p}=\sigma_{{Z_b^{\prime-}}p}=\sigma_{{Z_b^+}n},\,\,\,\,\sigma_{{Z_b^-}n}=\sigma_{{Z_b^{\prime-}}n}=
\sigma_{{Z_b^+}p}.
\end{equation}
It is seen that the $Z_b(10610)^{\pm}$ and $Z_b(10650)^{\pm}$ states as $B{\bar B}^*$ and $B^*{\bar B}^*$ molecules have the largest absorption cross sections. Hence, they are expected to be more easily absorbed in this case in a nuclear medium than their other intrinsic configurations. It should be also noted that according to Eqs. (3)--(5), (13), (28)--(30) for nuclei, like $^{12}_{6}$C, $^{40}_{20}$Ca, with the same local proton $\rho_p(r)$ and neutron $\rho_n(r)$ densities the $Z_b(10610)^+$ and $Z_b(10610)^-$ as well as $Z_b(10650)^+$ and $Z_b(10650)^-$
total and differential (defined below) production cross sections are the same.
\begin{figure}[!h]
\begin{center}
\includegraphics[width=15.0cm]{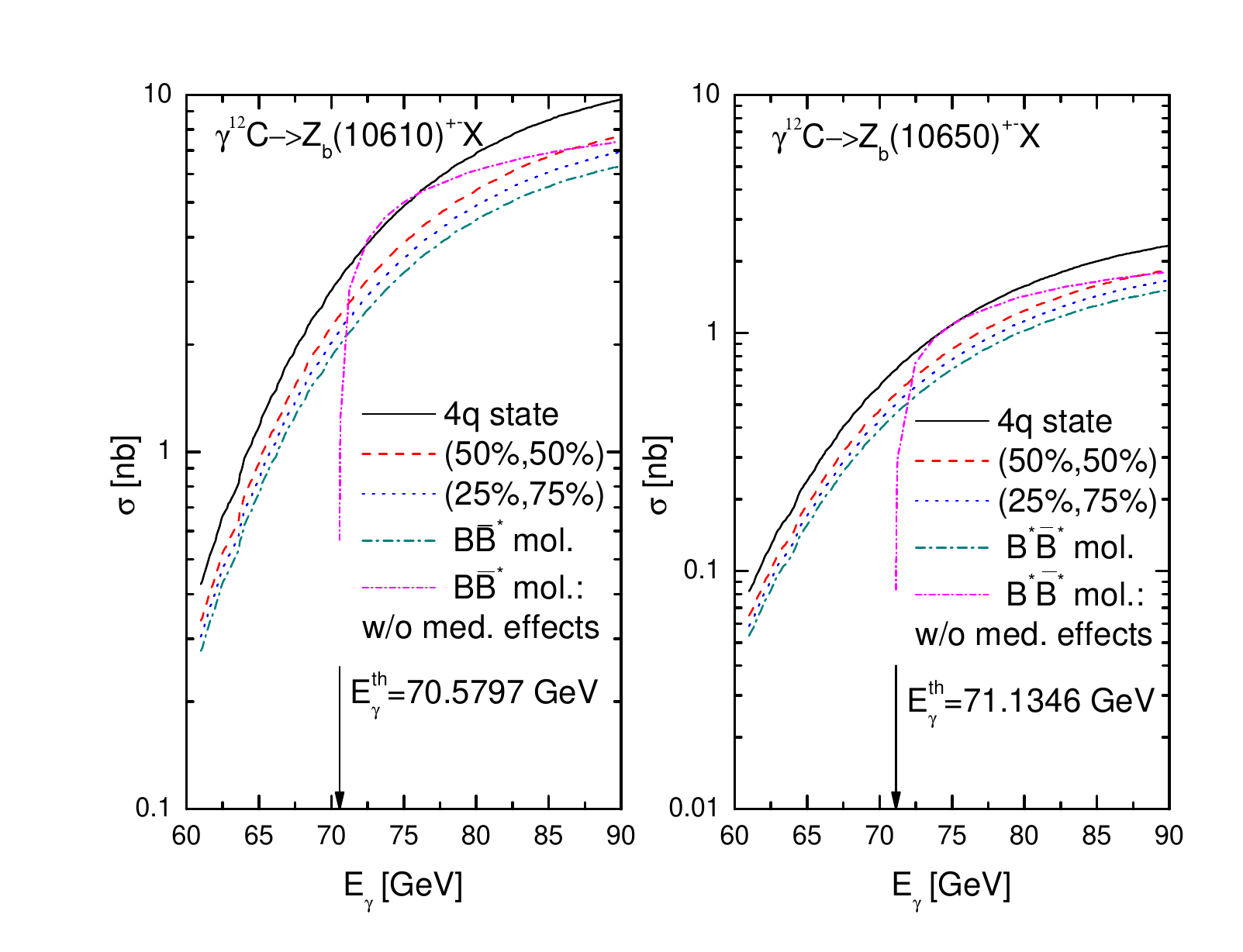}
\vspace*{-2mm} \caption{(Color online.) Excitation functions for the production of $Z_b(10610)^{\pm}$ (left panel) and
$Z_b(10650)^{\pm}$ (right panel) mesons off $^{12}$C from the direct processes (1), (2)
proceeding on an off-shell target nucleons and on a free ones being at rest. The curves are calculations in
the scenarios, in which the $Z_b(10610)^{\pm}$ and $Z_b(10650)^{\pm}$ are treated as a purely compact four quark (4$q$) states, as a purely molecular states, or as a hybrid states: mixtures of the non-molecular (compact) and molecular (non-compact) components, in which there are 50\% of the 4$q$ components and 50\% molecular components,
as well as 25\% and 75\% of the nonmolecular and molecular states. The left and right arrows indicate the threshold energies for the $Z_b(10610)^{\pm}$ and $Z_b(10650)^{\pm}$ photoproduction on a free nucleons in reactions (1), (2).}
\label{void}
\end{center}
\end{figure}

We consider now the following additional integral observables that can be used to verify properties of the $Z_b(10610/10650)^{\pm}$ resonances. They are cross-section ratios, which are sensitive to the $Z_b(10610/10650)^{\pm}$--nucleon absorption cross sections (and, hence, to their intrinsic structures). On the other  hand, they are less sensitive than cross sections themselves to the theoretical uncertainties associated mainly with the experimentally unknown total cross sections of the elementary reactions (1), (2). The first two observables - the so-called $Z_b^+(Z_b^{\prime{+}})$ and $Z_b^-(Z_b^{\prime{-}})$ transparency ratios - are the ratios of the nuclear $Z_b^+(Z_b^{\prime{+}})$ and $Z_b^-(Z_b^{\prime{-}})$ photoproduction cross sections (3), (4) divided, respectively,
by $Z$ and $N$ times the same quantities on a free proton and neutron (cf. Ref. [66] and references herein):
\begin{equation}
S_A^+=\frac{\sigma_{{\gamma}A \to Z_b^+(Z_b^{\prime{+}})X}^{({\rm dir})}(E_{\gamma})}{Z~\sigma_{{\gamma}p \to Z_b^+(Z_b^{\prime{+}})n}(\sqrt{s(E_{\gamma})})},
\end{equation}
\begin{equation}
S_A^-=\frac{\sigma_{{\gamma}A \to Z_b^-(Z_b^{\prime{-}})X}^{({\rm dir})}(E_{\gamma})}{N~\sigma_{{\gamma}n \to Z_b^-(Z_b^{\prime{-}})p}(\sqrt{s(E_{\gamma})})}.
\end{equation}
The second two observables are the $Z_b^{\pm}(Z_b^{\prime{\pm}})$ transparency ratios $S_A^{\pm}$ normalized to a light
nucleus like $^{12}$C [66, 110]:
\begin{equation}
T_A^+=\frac{S_A^{+}}{S_C^{+}}=\frac{6}{Z}\frac{\sigma_{{\gamma}A \to Z_b^+(Z_b^{\prime{+}})X}^{({\rm dir})}(E_{\gamma})}
{\sigma_{{\gamma}C \to Z_b^+(Z_b^{\prime{+}})X}^{({\rm dir})}(E_{\gamma})},
\end{equation}
\begin{equation}
T_A^-=\frac{S_A^{-}}{S_C^{-}}=\frac{6}{N}\frac{\sigma_{{\gamma}A \to Z_b^-(Z_b^{\prime{-}})X}^{({\rm dir})}(E_{\gamma})}
{\sigma_{{\gamma}C \to Z_b^-(Z_b^{\prime{-}})X}^{({\rm dir})}(E_{\gamma})}.
\end{equation}
It should be pointed out that the definition of the quantities $S_A^{\pm}$ and $T_A^{\pm}$ via Eqs. (31)--(34) implies that they should be considered only at above threshold photon energies. But since the right-hand sides of Eqs. (33)
and (34) are defined both above and below the respective thresholds, we will adopt them in calculating the transparency ratios $T_A^{\pm}$ also at subthreshold photon energies (cf. Fig. 8 given below).
\begin{figure}[!h]
\begin{center}
\includegraphics[width=15.0cm]{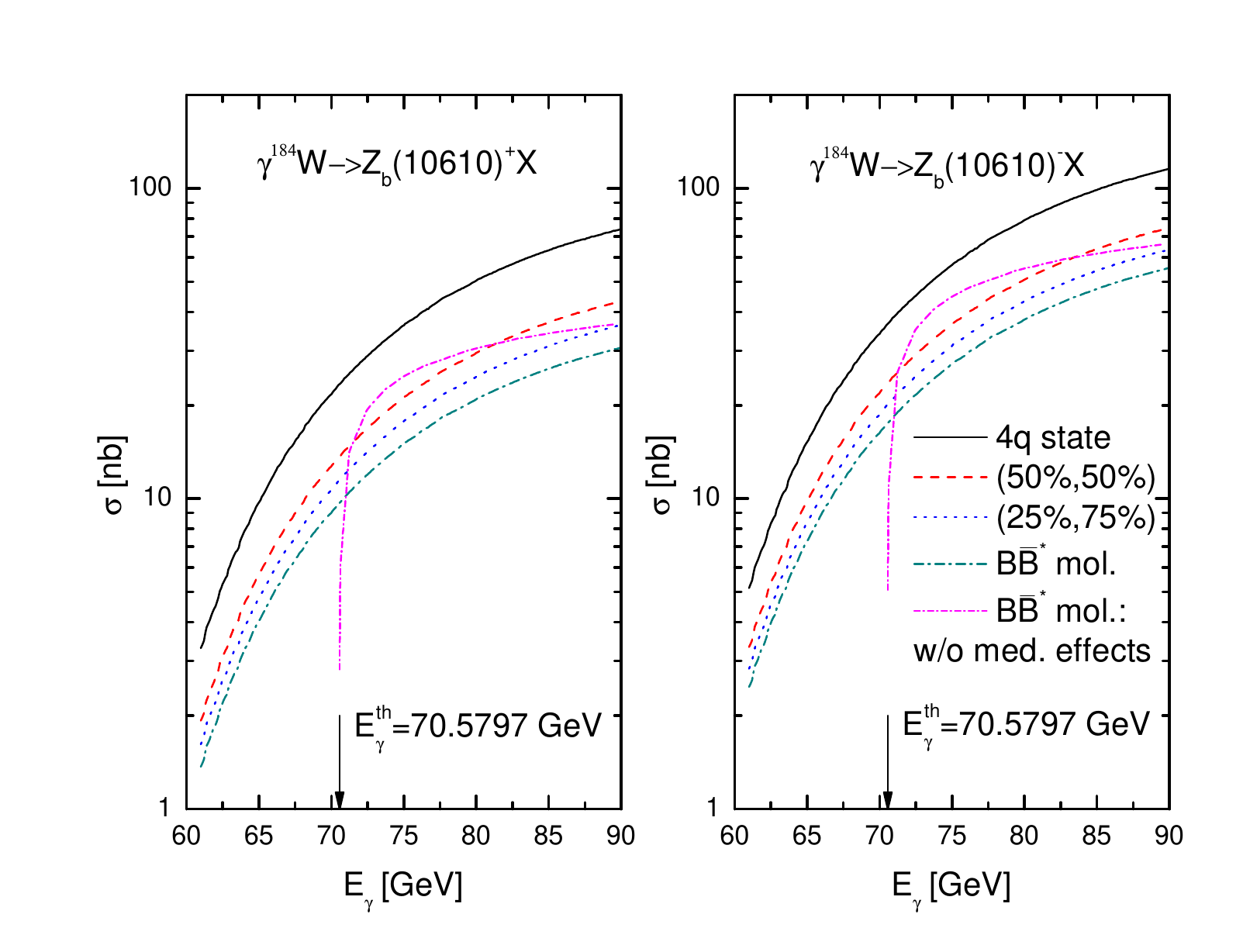}
\vspace*{-2mm} \caption{(Color online.) Excitation functions for the production of $Z_b(10610)^+$ (left panel)
and $Z_b(10610)^-$ (right panel) mesons off $^{184}$W , respectively, from the direct processes (1) and (2)
proceeding on an off-shell target nucleons and on a free ones being at rest. The curves and the arrows denote the same as in Fig. 1 for the $Z_b(10610)^{\pm}$.}
\label{void}
\end{center}
\end{figure}
\begin{figure}[!h]
\begin{center}
\includegraphics[width=15.0cm]{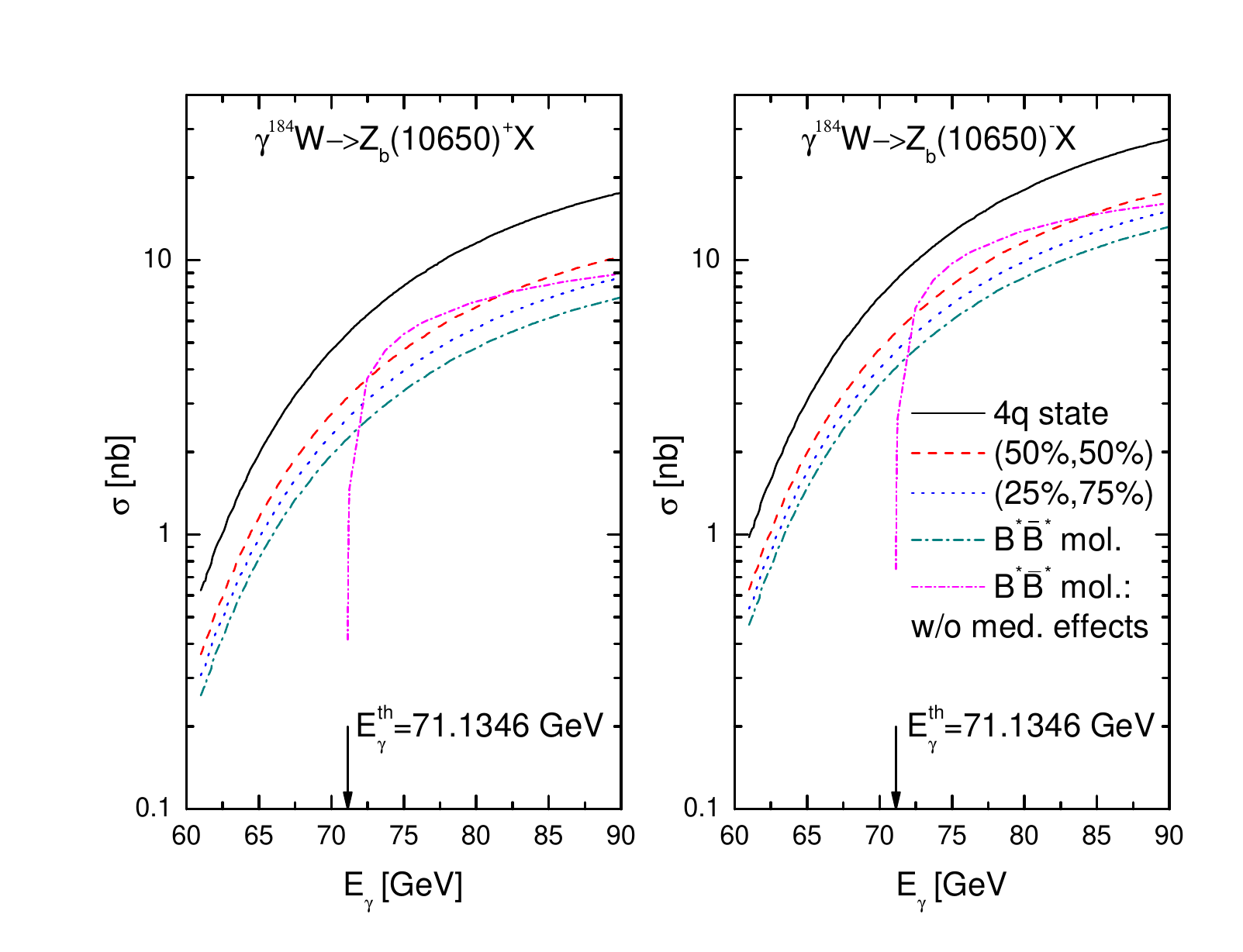}
\vspace*{-2mm} \caption{(Color online.) The same as in Fig. 2, but for the $Z_b(10650)^{\pm}$ mesons.}
\label{void}
\end{center}
\end{figure}

The information on the $Z_b^{\pm}(Z_b^{\prime{\pm}})$ intrinsic structures can also be extracted from the comparison of the measured and calculated their momentum distributions from nuclei in the photon energy range of interest.
Therefore, we consider now the momentum-dependent inclusive differential cross sections for their production
with momentum $p_{Z_b^{\pm}}(p_{Z_b^{\prime{\pm}}})$ on nuclei from the elementary production channels (1), (2).
Accounting for the fact that the $Z_b^{\pm}(Z_b^{\prime{\pm}})$ mesons move in the nucleus essentially forward in the laboratory system
\footnote{$^)$Thus, for example, the maximum angle of the $Z_b(10610)^+$ production on a free target proton at rest in
reaction (1) is about 0.7$^{\circ}$ at photon energy of 75 GeV.}$^)$,
we will calculate the $Z_b^{\pm}(Z_b^{\prime{\pm}})$ momentum distributions from the considered target nuclei
for the laboratory solid angles
${\Delta}{\bf \Omega}_{Z_b^{\pm}(Z_b^{\prime{\pm}})}$ = $0^{\circ} \le \theta_{Z_b^{\pm}(Z_b^{\prime{\pm}})} \le 5^{\circ}$, and $0 \le \varphi_{Z_b^{\pm}(Z_b^{\prime{\pm}})} \le 2{\pi}$. Then, according to Eqs. (3)--(5)
and Ref. [66], we represent these distributions as follows:
\begin{equation}
\frac{d\sigma_{{\gamma}A\to {Z_b^+}X}^{({\rm dir})}
(p_{\gamma},p_{Z_b^+})}{dp_{Z_b^+}}=
2{\pi}I_{V}[A,\rho_p,\sigma_{Z_b^+p},\sigma_{Z_b^+n}]
\int\limits_{\cos5^{\circ}}^{1}d\cos{{\theta_{Z_b^+}}}
\left<\frac{d\sigma_{{\gamma}p\to {Z_b^+}{n}}(p_{\gamma},
p_{Z_b^+},\theta_{Z_b^+})}{dp_{Z_b^+}d{\bf \Omega}_{Z_b^+}}\right>_A,
\end{equation}
\begin{equation}
\frac{d\sigma_{{\gamma}A\to {Z_b^-}X}^{({\rm dir})}
(p_{\gamma},p_{Z_b^-})}{dp_{Z_b^-}}=2{\pi}I_{V}[A,\rho_n,\sigma_{Z_b^-p},\sigma_{Z_b^-n}]
\int\limits_{\cos5^{\circ}}^{1}d\cos{{\theta_{Z_b^-}}}
\left<\frac{d\sigma_{{\gamma}n\to {Z_b^-}{p}}(p_{\gamma},
p_{Z_b^-},\theta_{Z_b^-})}{dp_{Z_b^-}d{\bf \Omega}_{Z_b^-}}\right>_A.
\end{equation}
Here,
$\left<\frac{d\sigma_{{\gamma}p \to Z_b^+n}(p_{\gamma},
p_{Z_b^{+}},\theta_{Z_b^{+}})}{dp_{Z_b^{+}}d{\bf \Omega}_{Z_b^{+}}}\right>_A$ and
$\left<\frac{d\sigma_{{\gamma}n \to Z_b^-p}(p_{\gamma},
p_{Z_b^{-}},\theta_{Z_b^{-}})}{dp_{Z_b^{-}}d{\bf \Omega}_{Z_b^{-}}}\right>_A$
are the off-shell differential cross sections for production of $Z_b^{+}$ and $Z_b^-$ mesons
with momenta ${\bf p}_{Z_b^{+}}$ and ${\bf p}_{Z_b^{-}}$ in the processes (1) and (2),
averaged over the Fermi motion and binding energy of the target nucleons.
These cross sections can be expressed by Eqs. (28), (31)--(39) from Ref. [111], in which one needs to make the
substitution: $({\gamma}p \to \Upsilon(1S)p) \to ({\gamma}p \to Z_b^+n)$, $\Upsilon(1S) \to Z_b^+$ and
$({\gamma}p \to \Upsilon(1S)p) \to ({\gamma}n \to Z_b^-p)$, $\Upsilon(1S) \to Z_b^-$.
The differential cross sections for the production of $Z_b^{\prime{+}}$ and
$Z_b^{\prime{-}}$ mesons on nuclei from the processes (1) and (2) can be expressed by Eqs. (35) and (36),
correspondingly, in which one needs to make the substitutions: $Z_b^{+} \to Z_b^{\prime{+}}$ and $Z_b^{-} \to Z_b^{\prime{-}}$.
To calculate the c.m. $Z_b^{\pm}$ and $Z_b^{\prime{\pm}}$ angular distributions in reactions (1), (2)
(cf. Eq. (34) from Ref. [111]) one needs to know their exponential $t$-slope parameters $b_{Z_b^{\pm}}$ and
$b_{Z_b^{\prime{\pm}}}$ in the energy region of interest. We adopt for both these parameters the value of
2.25 GeV$^{-2}$, corresponding [111] to the $\Upsilon(1S)$ slope parameter $b_{\Upsilon(1S)}$ in the reaction
${\gamma}p \to \Upsilon(1S)p$ at incident photon energy of 75 GeV.
\begin{figure}[!h]
\begin{center}
\includegraphics[width=15.0cm]{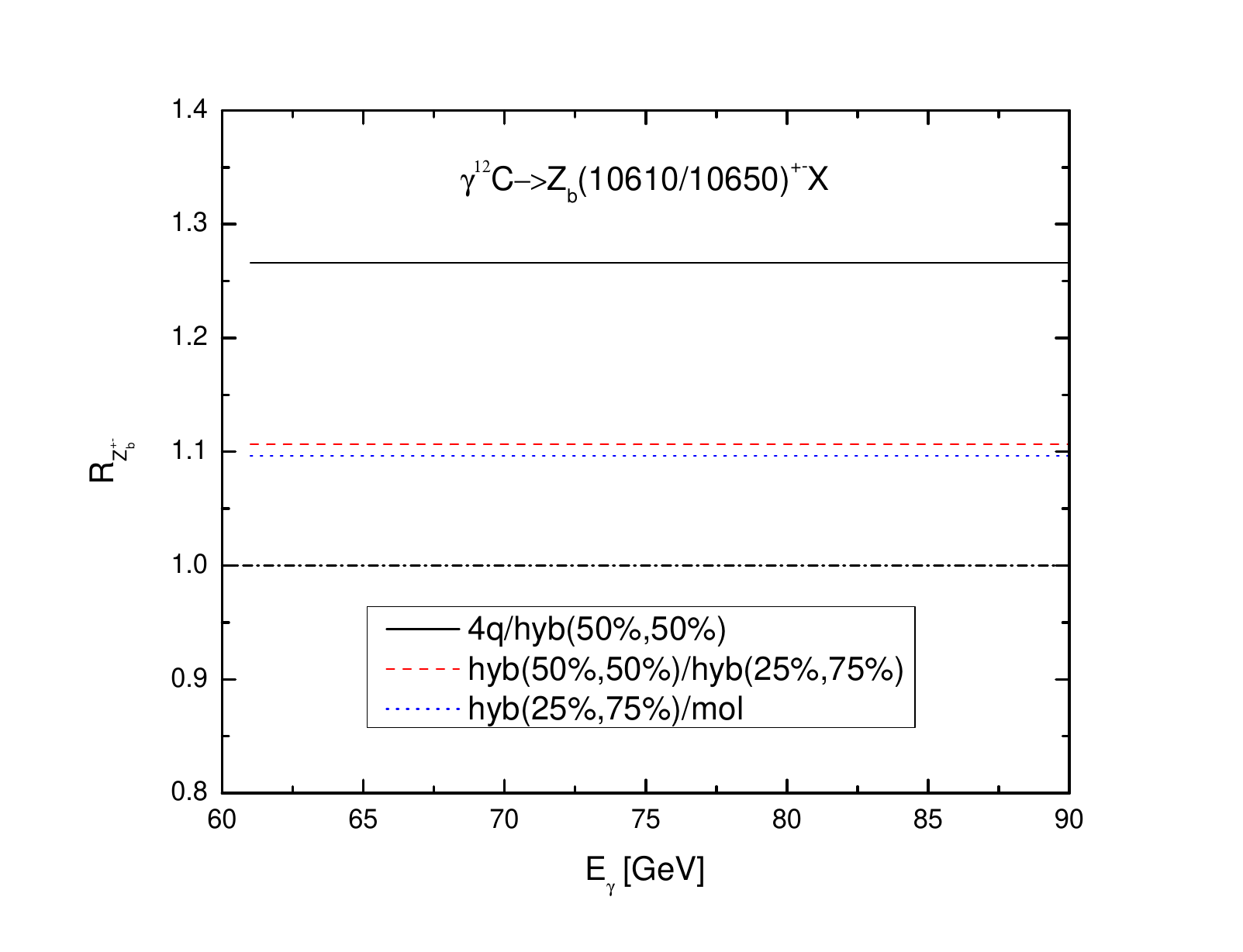}
\vspace*{-2mm} \caption{(Color online.) Ratios between the $Z_b(10610/10650)^{\pm}$ production cross sections on $^{12}$C, shown in Fig. 1 and calculated in the compact tetraquark scenarios and in the hybrid pictures for them, in which there are 50\% of the 4$q$ components and 50\% molecular components;
in the hybrid pictures for them, in which there are 50\% of the 4$q$ components and 50\% molecular components, as well as 25\% and 75\% of the nonmolecular and molecular states; in the hybrid pictures for them, in which there are 25\% of the 4$q$ components and 75\% molecular components, as well as in the molecular scenarios for an off-shell target nucleons, as functions of photon energy.}
\label{void}
\end{center}
\end{figure}
\begin{figure}[!h]
\begin{center}
\includegraphics[width=15.0cm]{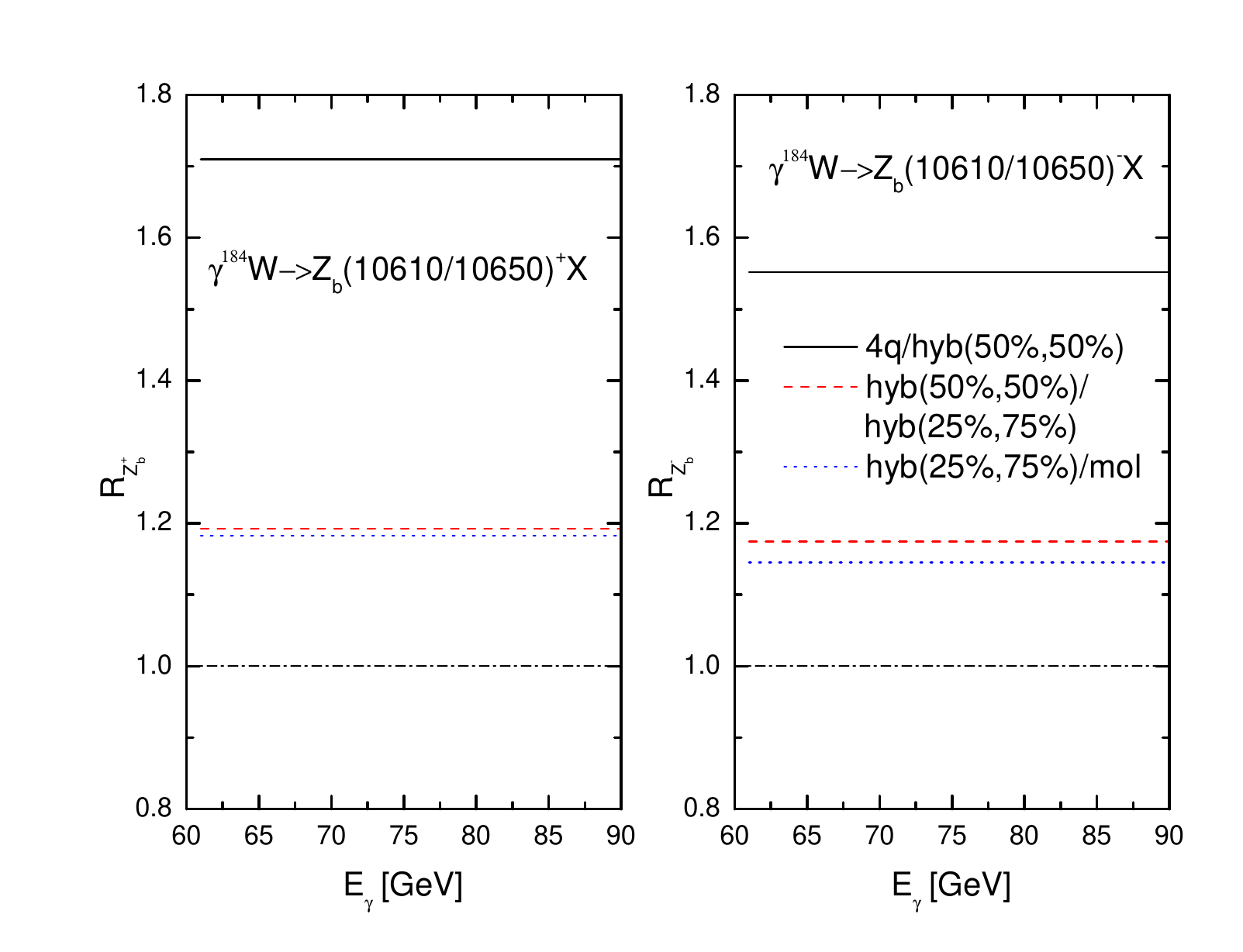}
\vspace*{-2mm} \caption{(Color online.) Ratios between the $Z_b(10610/10650)^{+}$ (left panel) and between the $Z_b(10610/10650)^-$ (right panel) production cross sections on $^{184}$W, shown in Figs. 2, 3, as functions of photon energy. The curves denote the same as in Fig. 4.}
\label{void}
\end{center}
\end{figure}
\begin{figure}[!h]
\begin{center}
\includegraphics[width=15.0cm]{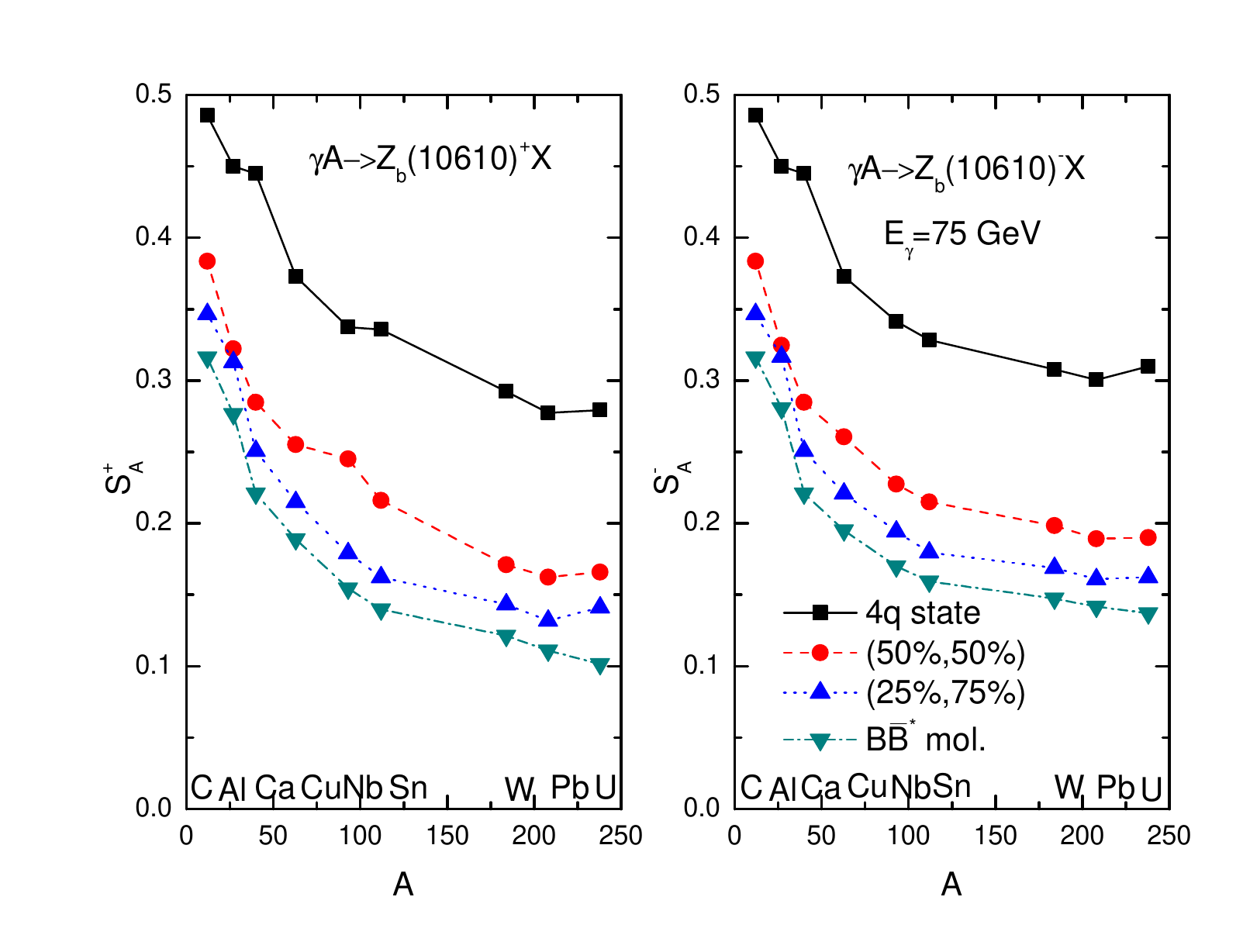}
\vspace*{-2mm} \caption{(Color online.) Transparency ratios $S_A$ for the $Z_b(10610)^+$ (left panel) and $Z_b(10610)^-$ (right panel) mesons, respectively, from the direct processes (1) and (2) proceeding on an off-shell target nucleons at incident photon energy of 75 GeV in the laboratory system as functions of the nuclear mass number $A$ in the considered theoretical pictures describing their intrinsic structures.
The lines are to guide the eyes.}
\label{void}
\end{center}
\end{figure}
\begin{figure}[!h]
\begin{center}
\includegraphics[width=15.0cm]{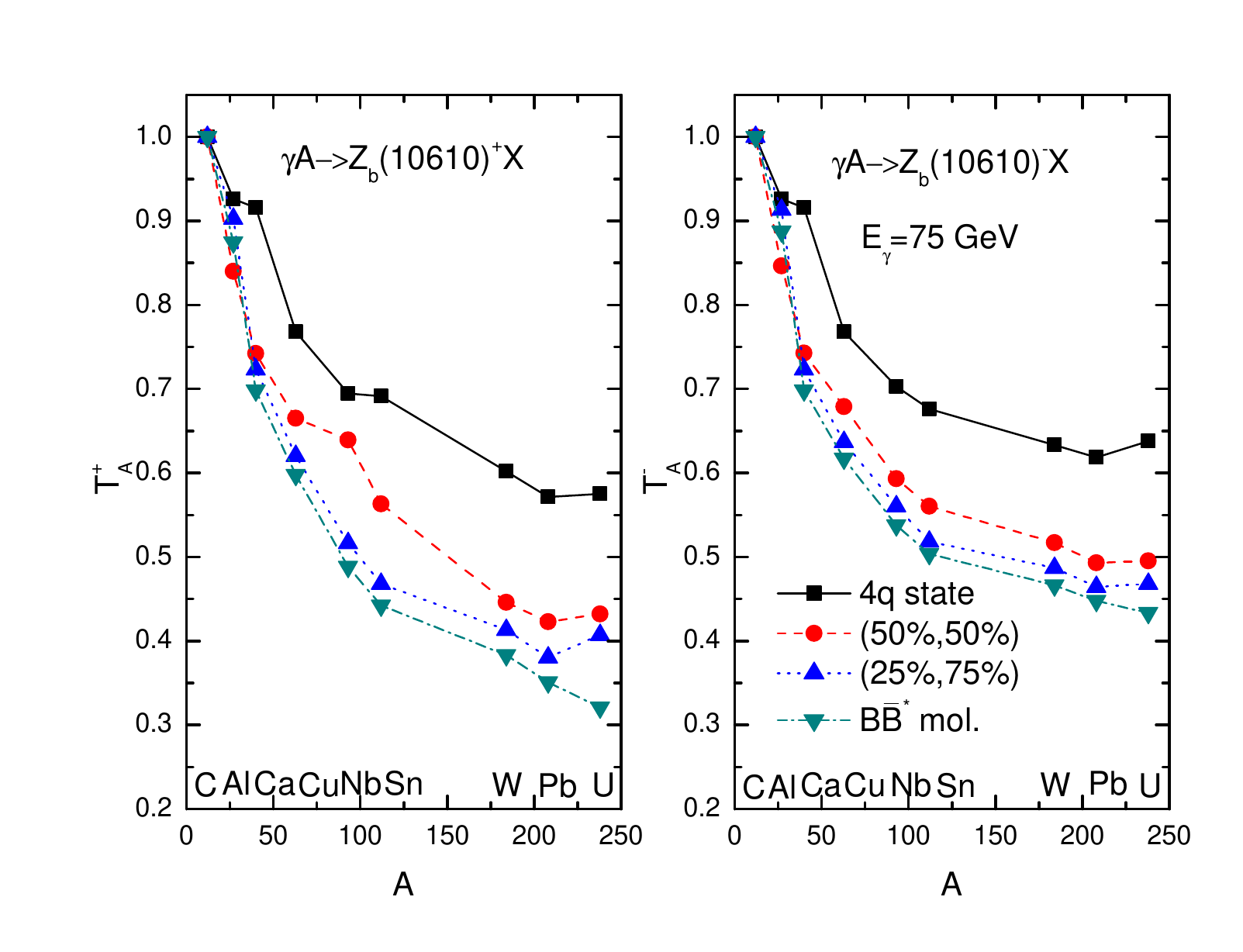}
\vspace*{-2mm} \caption{(Color online.) The same as in Fig. 6, but for the transparency ratios $T_A$.}
\label{void}
\end{center}
\end{figure}
\begin{figure}[!h]
\begin{center}
\includegraphics[width=15.0cm]{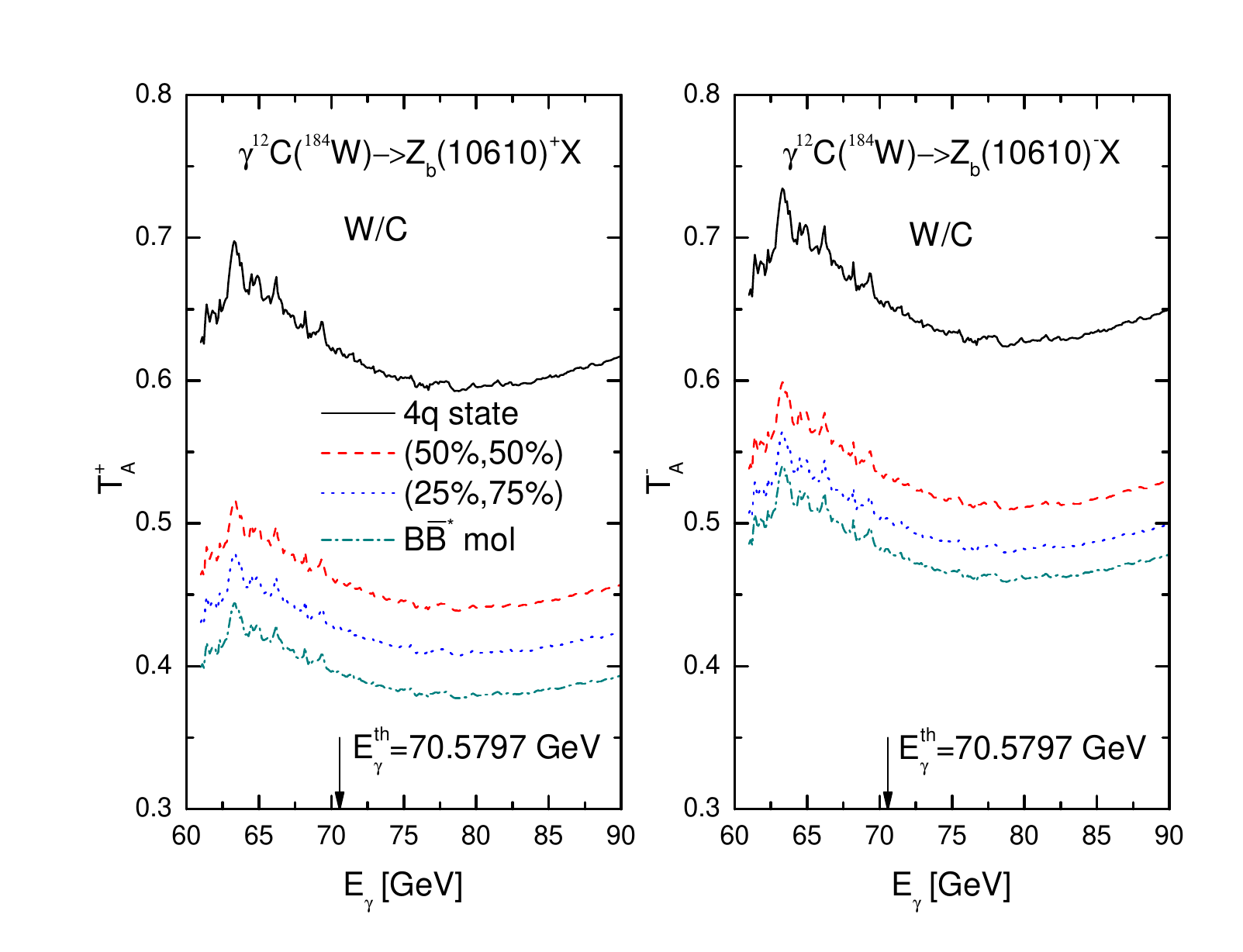}
\vspace*{-2mm} \caption{(Color online.) Transparency ratios $T_A$ for the $Z_b(10610)^+$ (left panel) and
$Z_b(10610)^-$ (right panel) mesons, respectively, from the direct processes (1) and (2) proceeding on an off-shell target nucleons as functions of the incident photon energy for combination $^{184}$W/$^{12}$C in the
considered theoretical pictures describing their intrinsic structures. The arrows indicate the threshold energy for the $Z_b(10610)^+$ and $Z_b(10610)^-$ photoproduction on a free nucleons in reactions (1) and (2), correspondingly.}
\label{void}
\end{center}
\end{figure}
\begin{figure}[!h]
\begin{center}
\includegraphics[width=15.0cm]{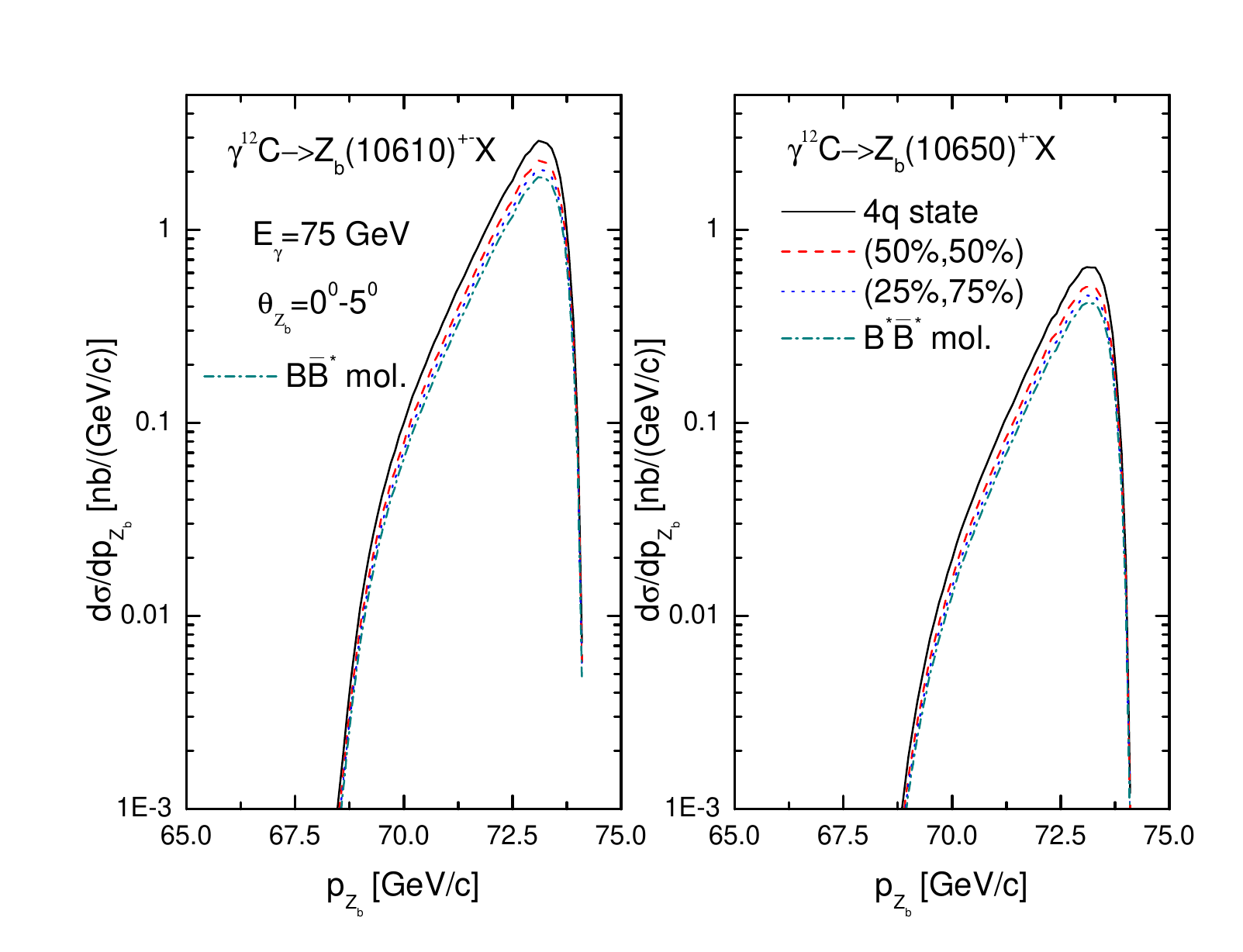}
\vspace*{-2mm} \caption{(Color online.) Momentum differential cross sections for the production of
$Z_b(10610)^{\pm}$ (left panel) and $Z_b(10650)^{\pm}$ mesons from the direct reactions
(1), (2) proceeding on an off-shell target nucleons in the laboratory polar angular range of 0$^{\circ}$--5$^{\circ}$ in the interaction of photons with energy of $E_{\gamma}=$ 75 GeV with $^{12}$C target nucleus in the considered theoretical pictures describing their intrinsic structures.}
\label{void}
\end{center}
\end{figure}
\begin{figure}[!h]
\begin{center}
\includegraphics[width=15.0cm]{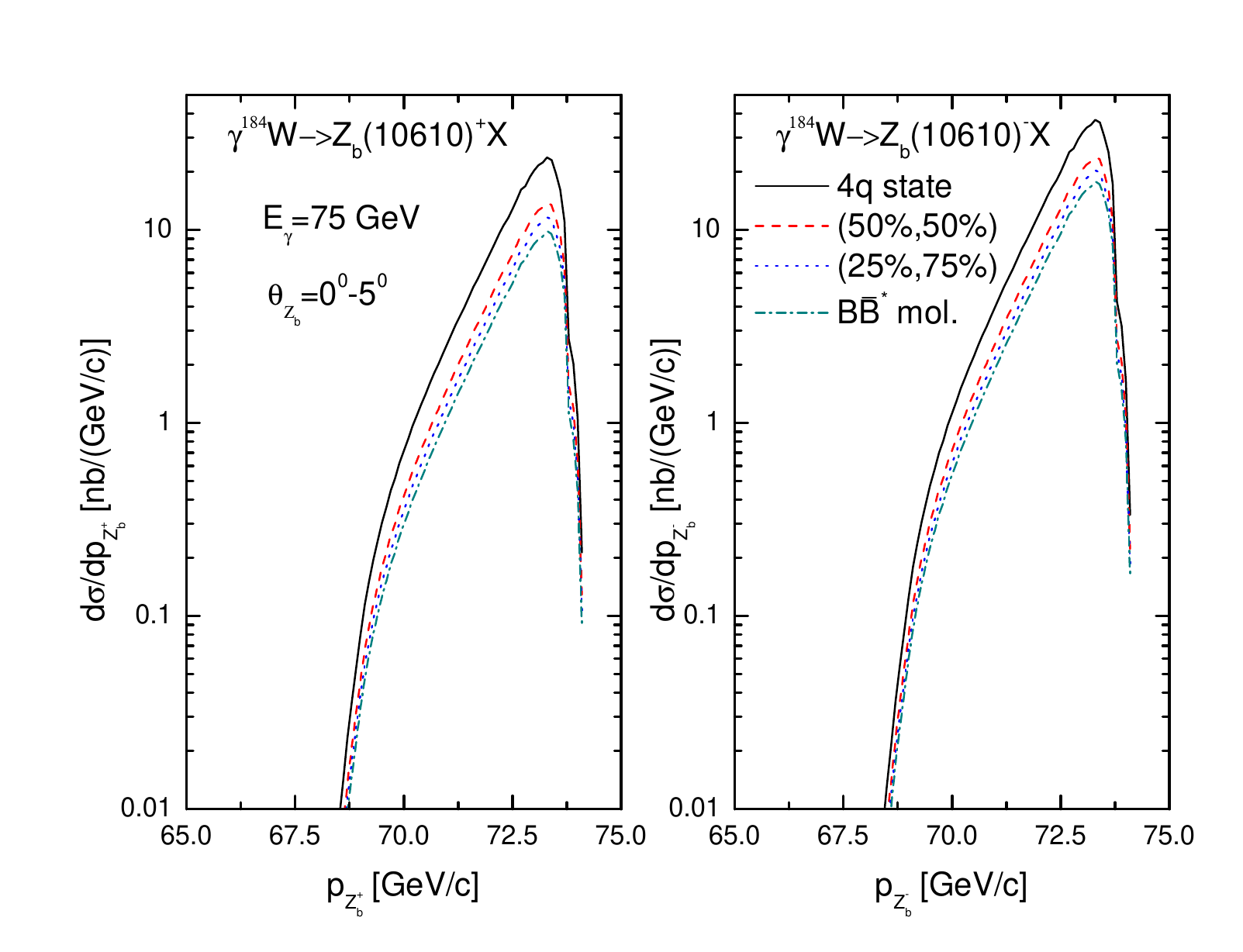}
\vspace*{-2mm} \caption{(Color online.) The same as in Fig. 9, but only for the $^{184}$W target nucleus
and for the $Z_b(10610)^+$ (left panel) and $Z_b(10610)^-$ (right panel) mesons.}
\label{void}
\end{center}
\end{figure}
\begin{figure}[!h]
\begin{center}
\includegraphics[width=15.0cm]{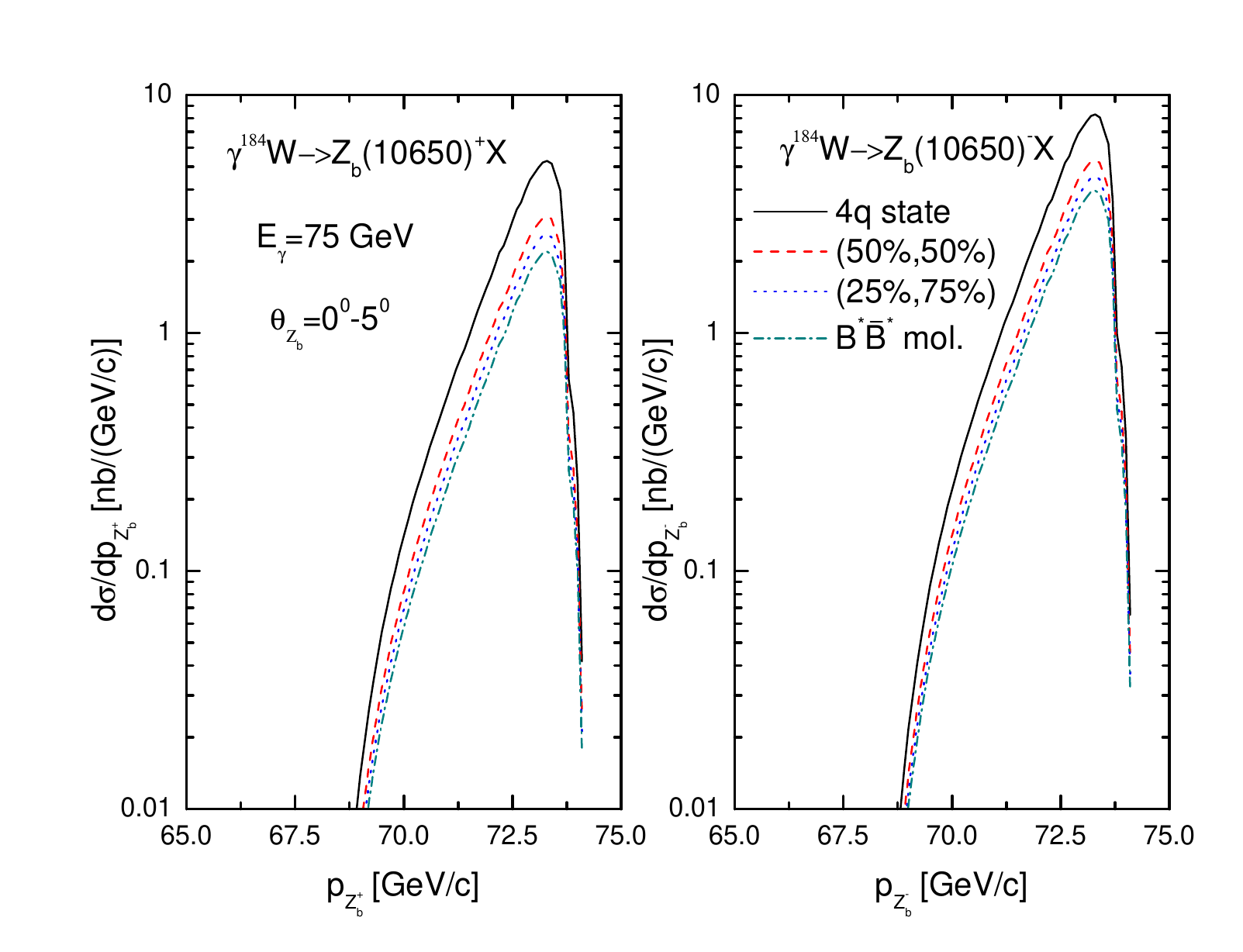}
\vspace*{-2mm} \caption{(Color online.) The same as in Fig. 10, but only for the
$Z_b(10650)^+$ (left panel) and $Z_b(10650)^-$ (right panel) mesons.}
\label{void}
\end{center}
\end{figure}
\begin{figure}[!h]
\begin{center}
\includegraphics[width=15.0cm]{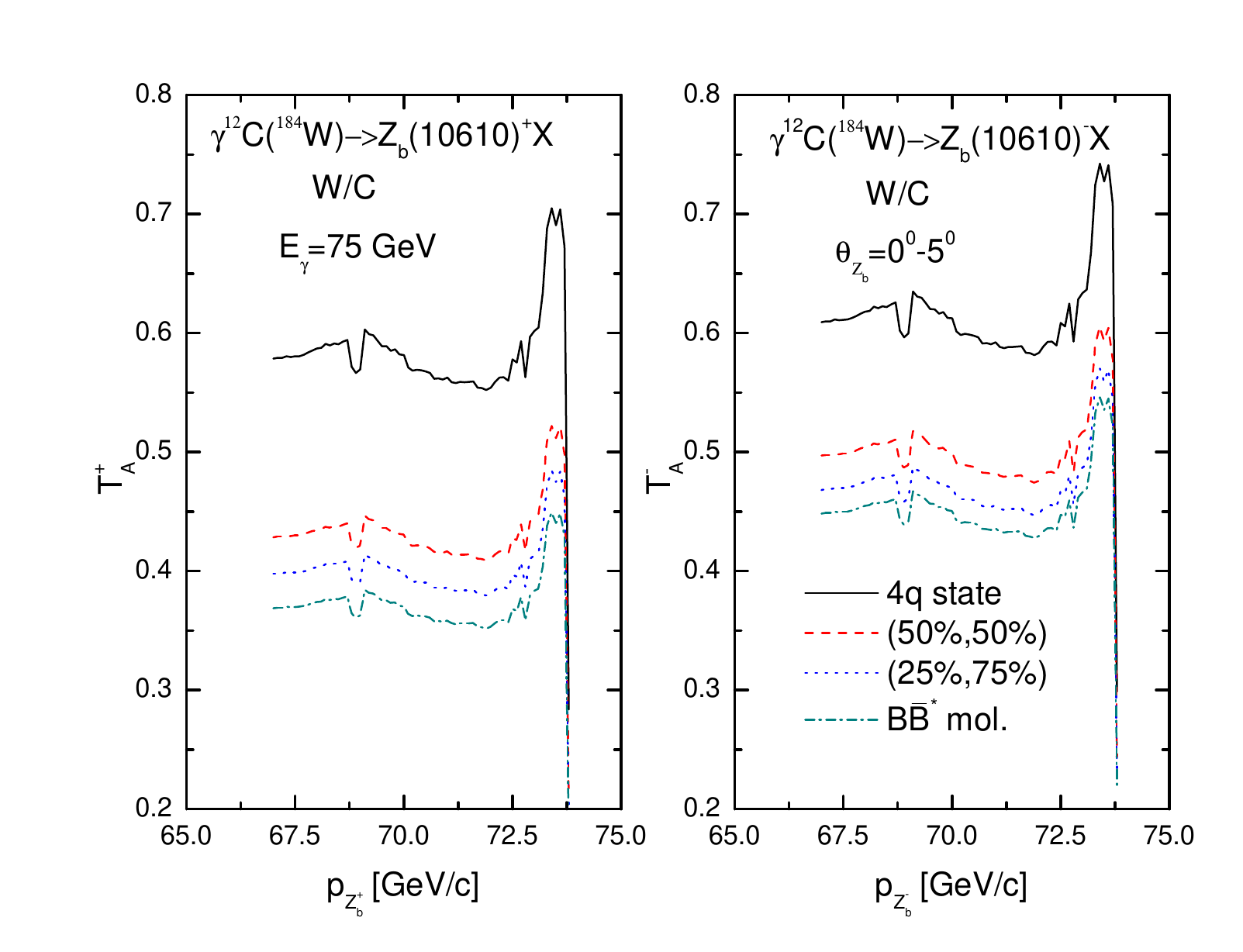}
\vspace*{-2mm} \caption{(Color online.) Transparency ratios $T_A$ for the $Z_b(10610)^+$ (left panel) and
$Z_b(10610)^-$ (right panel) mesons, respectively, from the direct processes (1) and (2) proceeding on an off-shell target nucleons as functions of their laboratory momenta for incident photon energy of 75 GeV for combination $^{184}$W/$^{12}$C, calculated in the laboratory polar angular range of 0$^{\circ}$--5$^{\circ}$
in the considered theoretical pictures describing their intrinsic structures.}
\label{void}
\end{center}
\end{figure}

\section*{3. Numerical results and discussion}

\hspace{1.5cm} The excitation functions for production of $Z_b(10610)^{\pm}$ and $Z_b(10650)^{\pm}$ mesons on $^{12}$C and on $^{184}$W target nuclei, calculated in line with Eqs. (3), (4) for four adopted options for
the $Z_b^{\pm}$(${Z_b^{\prime{\pm}}}$) absorption cross sections $\sigma_{Z_b^{\pm}({Z_b^{\prime{\pm}}})p(n)}$ in
nuclei (cf. Eqs. (28)--(30)) as well as for an off-shell target nucleons and for a free ones being at rest, are shown in Figs. 1 and 2, 3, respectively. It is seen that the difference between calculations with and without accounting for the target nucleon Fermi motion (between magenta short-dashed-dotted and dark cyan dashed-dotted curves)
is sufficiently small at above threshold photon energies, while at subthreshold photon  energies its impact on the $Z_b(10610)^{\pm}$ and $Z_b(10650)^{\pm}$ yields is essential. It is also seen that the absolute values of these yields have at well above threshold photon energies $\sim$ 80--85 GeV a measurable  strengths $\sim$ 10 nb for $Z_b(10610)^{\pm}$ and $\sim$ 2 nb for $Z_b(10650)^{\pm}$ for carbon as well as $\sim$ 100 nb for $Z_b(10610)^{\pm}$ and
$\sim$ 20 nb for $Z_b(10650)^{\pm}$ for tungsten.
For both considered target nuclei and for all photon energies studied the above yields reveal a definite sensitivity to the $Z_b(10610)^{\pm}$ and $Z_b(10650)^{\pm}$ intrinsic configurations encoded in their absorption cross sections (28)--(30). Thus, for the heavy target nucleus $^{184}$W there are a well distinguishable and experimentally measurable differences between the calculations corresponding to these configurations, namely: between the calculations corresponding to the pure $B{\bar B}^*$ and $B^*{\bar B}^*$ molecular states and hybrid states with the non-molecular and molecular probabilities of 25\% and 75\%, to the
hybrid states with the non-molecular and molecular probabilities of 25\% and 75\% and hybrid states with the non-molecular and molecular probabilities of 50\% and 50\%, to the hybrid states with the non-molecular and molecular probabilities of 50\% and 50\% and pure compact tetraquarks states. They, correspondingly, are $\sim$ 20, 20, 70\% for the $Z_b(10610/10650)^+$ and $\sim$ 15, 20, 55\% for the $Z_b(10610/10650)^-$. For the light target nucleus $^{12}$C, the sensitivity of the $Z_b(10610)^{\pm}$ and $Z_b(10650)^{\pm}$ production cross sections to their inner structures becomes lower and the same differences as above become smaller and almost insignificant. They, respectively, are $\sim$ 10, 10, 27\% both for the $Z_b(10610/10650)^+$ and for the $Z_b(10610/10650)^-$. Nevertheless, one might expect to measure them as well in the future high-precision photoproduction experiments at the proposed electron-ion colliders [62--65]
\footnote{$^)$It should be pointed out that the differences between the calculations corresponding to the pure $B{\bar B}^*$ and $B^*{\bar B}^*$ molecular states and hybrid states with the non-molecular and molecular probabilities of 50\% and 50\% are $\sim$ 40 and 35\% for $^{184}$W and $\sim$ 21\% for $^{12}$C for the $Z_b(10610/10650)^+$ and for the $Z_b(10610/10650)^-$. In view of the above, this means that the precise $Z_b$ and $Z_b^{\prime}$ photoproduction data, obtained in such experiments, should definitely help to discriminate, at least, between a compact tetraquark,
compact tetraquark-molecule mixtures with the compact tetraquark and molecular probabilities $\sim$ 50\% and 50\% and a pure molecular configurations of the $Z_b$ and $Z_b^{\prime}$ exotic hadrons.}$^)$.
To motivate the measurements at these colliders also in electroproduction, it is desirable to estimate, for example, the $Z_b(10610/10650)^{+}$ production rates (the event numbers) in the $e^{12}$C and $e^{184}$W reactions in a one-year run.
For this purpose, at first, we translate the $Z_b(10610/10650)^+$ photoproduction total cross sections, reported above,
namely: 10 and 100 nb/2 and 20 nb into the expected total cross sections of the sequences [21]
${e}^{12}{\rm C}(^{184}{\rm W}) \to e^{\prime}{Z_b(10610/10650)^+}X$, $Z_b(10610/10650)^+ \to \Upsilon(2S)\pi^+ \to
{\mu^+}{\mu^-}\pi^+$ by multiplying them on the appropriate branching ratios
$Br[Z_b(10610)^+ \to \Upsilon(2S)\pi^+]\approx$ 3.6\%, $Br[Z_b(10650)^+ \to \Upsilon(2S)\pi^+]\approx$ 1.4\%,
$Br[\Upsilon(2S) \to {\mu^+}{\mu^-}]\approx$ 1.93\% [95] and by the factor 10$^{-2}$ to account for the fact that approximately two orders of magnitude smaller cross sections are anticipated for the electroproduction compared to the above photon production [112]. Then, to estimate the total numbers of the $Z_b(10610/10650)^+$ events in a one-year run at EicC, one needs to multiply the latter "electroproduction" total cross sections
on the carbon and tungsten target nuclei by the integrated luminosity of 60 fb$^{-1}$ [65, 113]
as well as by the detection efficiency. With a realistic 50\% detection efficiency [65], we estimate about of 2084 and 20840 events per year for the $Z_b(10610)^+$ signal in the cases of the $^{12}$C and $^{184}$W target nuclei, respectively. The respective event numbers for $Z_b(10610)^+$ at EIC are about of 10420 and 104200 with an integrated luminosity of 300 fb$^{-1}$ [63, 113]. The analogous event numbers for the $Z_b(10650)^+$ signals at EicC and EIC are
162 and 1620 and 810 and 8100. We see that the observation of the $Z_b(10610/10650)^+$ mesons is quite optimistic at EicC and especially at EIC also with the $eA$ reactions. Thus, the electron-ion colliders EicC and EIC provide a good platform to study the nature of the exotic tetraquark $Z_b(10610/10650)^{\pm}$ mesons as well as hidden-bottom pentaquarks
[111, 112, 114] in near-threshold reactions with electromagnetic probes.

To see more clearly the sensitivity of the total cross sections, presented in Figs. 1--3, to the in-medium absorption cross sections of the $Z_b(10610)^{\pm}$ and $Z_b(10650)^{\pm}$ mesons (or to their intrinsic configurations), we show in Figs. 4 and 5 on a linear scale the photon energy dependences of the ratios $R_{Z_b^{\pm}}$ between the cross sections presented, respectively, in Figs. 1 and 2, 3. It is seen that for the heavy $^{184}$W target nucleus
the sensitivity of the yields of $Z_b(10610)^{\pm}$ and $Z_b(10650)^{\pm}$ mesons to variations in their intrinsic structures is indeed significantly stronger than that for the light $^{12}$C target nucleus and it would definitely be possible for $^{184}$W nucleus to distinguish between the results corresponding to all considered choices for these structures at all photon beam energies. For the $^{12}$C nucleus we see a large enough distinctions only between the curves corresponding to the pure molecular configurations, hybrid configurations with the compact and non-compact probabilities of 50\% and 50\% and pure compact tetraquark configurations for the $Z_b(10610)^{\pm}$ and $Z_b(10650)^{\pm}$ mesons. This implies that their structures could be studied from the excitation function measurements at the future electron-ion colliders in inclusive near-threshold photonuclear reactions, once the $Z_b$ and $Z_b^{\prime}$ near-threshold proton (and neutron) target photoproduction cross sections will be experimentally known.

Additional information about the internal structure of the $Z_b(10610)^{\pm}$ mesons is contained in
Figs. 6 and 7, where the A-dependences of the transparency ratios $S_A^{\pm}$ and $T_A^{\pm}$ of $Z_b(10610)^{\pm}$ production from the direct processes (1), (2) in ${\gamma}A$
($A=$$^{12}$C, $^{27}$Al, $^{40}$Ca, $^{63}$Cu, $^{93}$Nb, $^{112}$Sn, $^{184}$W, $^{208}$Pb, and $^{238}$U) reactions
are presented. They have been calculated for photon energy of 75 GeV on the basis of Eqs. (31), (32) and (33), (34), respectively, and for the same values of the $Z_b(10610)^{\pm}$--nucleon absorption cross sections as those given above by Eqs. (28)--(30). One can see that the quantities $S_A^{\pm}$ and $T_A^{\pm}$ depend
strongly both on these cross sections (or on the considered interpretations of $Z_b(10610)^{\pm}$) and on the nuclear mass number $A$. Thus, they drop with increasing this number and reach values of the order of 0.1 and 0.3, correspondingly, for heavy nuclei like $^{208}$Pb and $^{238}$U in the $B{\bar B}^*$ molecular scenario -- a large deviation from unity which should be easily seen in a future experiments. Also, there are the following changes in them both for the $Z_b(10610)^{+}$ and for the $Z_b(10610)^{-}$ between calculations performed assuming for $Z_b(10610)^{\pm}$ pure molecular pictures and hybrid pictures with the non-molecular and molecular probabilities of 25\% and 75\%,
hybrid states with these probabilities of 25\% and 75\% and hybrid states with the above probabilities of 50\% and 50\% and hybrid states with the non-molecular and molecular probabilities of 50\% and 50\% and compact tetraquark states:
they, respectively, are $\sim$ 15, 15 and 50\% for relatively "light" nuclei ($^{27}$Al,$^{40}$Ca), $\sim$ 15, 30 and 50\% for the medium-mass ($^{93}$Nb,$^{112}$Sn) nuclei and $\sim$ 20, 20 and 70\% for heavy ($^{184}$W,$^{238}$U) target nuclei for the $S_A^{\pm}$ observables. For the $T_A^{\pm}$ observables the analogous changes are about 3, 5 and 16\%, 5, 14 and 18\% and 7, 7 and 30\%, respectively. They are smaller than those in previous case and are experimentally distinguishable only between calculations corresponding to the compact tetraquark and hybrid + molecular interpretations of the exotic $Z_b(10610)^{\pm}$  states. On the other hand, the differences between calculations corresponding to the
hybrid picture of these states with the non-molecular and molecular probabilities of 50\% and 50\% and pure molecular
picture are sizeable and also experimentally measurable, they are $\sim$ 20\% in the range of medium and large A.
Therefore, in line with the preceding, we conclude that the observation of the A-dependences of the transparency ratios $S_A^{\pm}$ and $T_A^{\pm}$, at least, for medium and large mass numbers $A$ in the future experiments at electron-ion colliders would also allow to elucidate the nature of the exotic QCD $Z_b(10610)^{\pm}$ states. An analogous conclusions can be also made, as our calculations have shown, for the $Z_b(10650)^{\pm}$ states. In the context of this,
the latter relative observables ($T_A^{\pm}$) are even more favorable than the former ones ($S_A^{\pm}$), since the theoretical uncertainties associated with the experimentally unknown elementary production cross sections substantially cancel out in them [109].

To provide some additional guidance for future experiments we show in Fig. 8 the photon energy dependences of the transparency ratios $T_A^{\pm}$ for $Z_b(10610)^{\pm}$ mesons for the $^{184}$W/$^{12}$C combination calculated in line with Eqs. (33), (34) using results presented in Figs. 1 and 2, respectively. It is seen that they can also be used for discriminating between possible scenarios for their inner configurations.

In addition to the $Z_b$, $Z_b^{\prime}$ integral observables, considered above, we discuss now their differential
observables which show the kinematic properties of these resonances and also provide a certain guidance for their future experimental detection. The absolute momentum distributions of the $Z_b(10610/10650)^{\pm}$ mesons from direct processes
(1), (2) in the ${\gamma}^{12}$C and ${\gamma}^{184}$W reactions, calculated on the basis of Eqs. (35), (36) for
adopted scenarios for their inner structures for laboratory polar angles of 0$^{\circ}$--5$^{\circ}$
and for photon energy of 75 GeV, are shown, respectively, in Figs. 9 and 10, 11. One can see that the absolute values of the distributions of the $Z_b(10610)^{\pm}$ mesons have a well measurable strengths $\sim$ 3 nb/(GeV/c) and 30 nb/(GeV/c) at central momenta around of 73 GeV/c in the cases of the $^{12}$C and $^{184}$W target nuclei, respectively. The analogous values for the $Z_b(10650)^{\pm}$ hadrons at these momenta are about of 0.5 nb/(GeV/c) and 5 nb/(GeV/c).
The differential cross sections considered also show a rather sizeable variations, especially for the heavy target nucleus $^{184}$W
\footnote{$^)$Which are similar to those shown in Figs. 2 and 3.}$^)$,
when going from the molecular to compact tetraquark assignments of the $Z_b(10610/10650)^{\pm}$ states.
Such behavior of them can also be used to clarify the nature of the exotic charged
bottomonium-like $Z_b(10610/10650)^{\pm}$ states from comparison the present model calculations with
data expected from experiments at the planned high-luminosity electron-ion colliders EIC and EicC in the US and China.

Finally, in Fig. 12 we show the "differential" transparency ratios $T_A^{\pm}$ for target combination W/C for
$Z_b(10610)^{\pm}$ mesons produced in elementary reactions (1), (2) at laboratory angles of 0$^{\circ}$--5$^{\circ}$
by 75 GeV photons as functions of their laboratory momenta. These ratios was calculated in line with Eqs. (33), (34) by adopting in them the differential cross sections presented in Figs. 9 and 10 instead of the total cross sections.
It is seen that the sensitivity of these ratios to different treatments of $Z_b(10610)^{\pm}$ mesons is similar to that available in Fig. 8. Therefore, they can be also used for discriminating between possible scenarios for their internal structures. Furthermore, the ratios have a weak momentum dependence practically at all momenta except of those belonging to the high-momentum region. This feature can be used as well to better limit these structures.
The above conclusions can be also made, as our calculations have shown, for the $Z_b(10650)^{\pm}$ states.

In the end, the absolute and relative observables considered in the present work can be useful to help determine
the $Z_b(10610/10650)^{\pm}$ intrinsic structures.

\section*{4. Epilogue}

\hspace{1.5cm}
In this paper, we have investigated the possibility to study the nature of the exotic charged bottomonium-like states $Z_b(10610)$ and $Z_b(10650)$ from their inclusive photoproduction off nuclei near the kinematic threshold  within the collision model based on the nuclear spectral function. The model accounts for $Z_b(10610)^{\pm}$ and $Z_b(10650)^{\pm}$ production in direct photon--nucleon interactions as well as four different scenarios for their intrinsic configurations: compact tetraquarks, molecules of the two open-beauty mesons and two mixtures of both of them for each of the $Z_b$ state. We have calculated within these scenarios the absolute and relative excitation functions on $^{12}$C and $^{184}$W nuclei at photon energies of 61--90 GeV, the absolute momentum differential cross sections and ratios of them for their production off these target nuclei at laboratory polar angles of 0$^{\circ}$--5$^{\circ}$ and for photon energy of 75 GeV as well as the A-dependences of the transparency ratios for the $Z_b(10610)^{\pm}$ mesons at photon energy of 75 GeV. We show that the absolute and relative observables considered reveal distinct sensitivity to the $Z_b(10610)^{\pm}$ and $Z_b(10650)^{\pm}$ internal structures. Therefore, they might be useful for determination of these structures from the comparison of them with the experimental data from the future high-precision experiments at the planned high-luminosity electron-ion colliders in the United States and China.
We hope that the results reported in the present work will stimulate the performing here such experiments and that
the theoretical framework developed in it will be also helpful in revealing the underlying structures of some another exotic hadrons.
\\

\end{document}